\documentclass[prb,twocolumn,amsmath,amssymb,superscriptaddress]{revtex4}
\usepackage{float}
\usepackage{amsmath}
\usepackage{amssymb}
\usepackage{amsfonts}
\usepackage{euscript}
\usepackage{enumerate}
\usepackage{hhline}
\usepackage{braket}
\usepackage{pslatex}
\usepackage{tabularx}
\usepackage[usenames,dvipsnames]{xcolor}
\usepackage{graphicx}
\usepackage{dcolumn}
\usepackage{bm}
\usepackage[sort&compress]{natbib}

\makeatletter
\renewcommand{\@biblabel}[1]{#1. }
\renewcommand{\@dotsep}{500}
\renewcommand{\@pnumwidth}{0em}
\renewcommand{\l@figure}[2]{
\@dottedtocline{1}{1.5em}{2em}{Figure #1}{}\vspace{15pt}}
\begin{document}

\title{Tunable quantum beat of single photons enabled by nonlinear nanophotonics}

\author{Qing Li}\email{qingli2@andrew.cmu.edu}
\affiliation{Physical Measurement Laboratory, National
	Institute of Standards and Technology, Gaithersburg, MD 20899,
	USA}
\affiliation{Maryland NanoCenter, University of Maryland,
	College Park, MD 20742, USA}
\affiliation{Electrical and Computer Engineering, Carnegie Mellon University, Pittsburgh, PA 15213, USA}
\author{Anshuman Singh}
\affiliation{Physical Measurement Laboratory, National
	Institute of Standards and Technology, Gaithersburg, MD 20899,
	USA}\affiliation{Maryland NanoCenter, University of Maryland,
	College Park, MD 20742, USA}
\author{Xiyuan Lu}
\affiliation{Physical Measurement Laboratory, National
	Institute of Standards and Technology, Gaithersburg, MD 20899,
	USA}\affiliation{Maryland NanoCenter, University of Maryland,
	College Park, MD 20742, USA}
\author{John Lawall}
\affiliation{Physical Measurement Laboratory, National Institute of Standards and Technology, Gaithersburg, MD 20899, USA}
\author{Varun Verma}
\affiliation{Physical Measurement Laboratory, National Institute of Standards and Technology, Boulder, CO 80305, USA}
\author{Richard Mirin}
\affiliation{Physical Measurement Laboratory, National Institute of Standards and Technology, Boulder, CO 80305, USA}
\author{Sae Woo Nam}
\affiliation{Physical Measurement Laboratory, National Institute of Standards and Technology, Boulder, CO 80305, USA}
\author{Kartik Srinivasan} \email{kartik.srinivasan@nist.gov}
\affiliation{Physical Measurement Laboratory, National
	Institute of Standards and Technology, Gaithersburg, MD 20899,
	USA}
\affiliation{Joint Quantum Institute, NIST/University of Maryland,
	College Park, MD 20742, USA}

\date{\today}

\begin{abstract}
{\noindent We demonstrate the tunable quantum beat of single photons through the co-development of core nonlinear nanophotonic technologies for frequency-domain manipulation of quantum states in a common physical platform. Spontaneous four-wave mixing in a nonlinear resonator is used to produce non-degenerate, quantum-correlated photon pairs.  One photon from each pair is then frequency shifted, without degradation of photon statistics, using four-wave mixing Bragg scattering in a second nonlinear resonator. Fine tuning of the applied frequency shift enables tunable quantum interference of the two photons as they are impinged on a beamsplitter, with an oscillating signature that depends on their frequency difference. Our work showcases the potential of nonlinear nanophotonic devices as a valuable resource for photonic quantum information science.}
\end{abstract}

\maketitle
\vspace{-0.1in}
\textit {Introduction}\textemdash Frequency-bin encoded states have attracted significant interest in quantum information processing due to the ease in realizing high-dimensional entangled states~\cite{lukens_frequency-encoded_2017,kues_-chip_2017,imany_50-ghz-spaced_2018}. This approach is particularly attractive in terms of resource scaling for chip-scale implementations. However, to fully explore such scalability, many physical resources need to be designed and reinvented using nanophotonics technology. For example, cross-modal coupling between frequency bins requires active elements that can efficiently provide controllable frequency shifts. Here, we demonstrate how silicon nanophotonics can support essential nonlinear nanophotonic technologies - quantum light generation and quantum frequency converter - to underpin such research.

In particular, we demonstrate the ability to flexibly tailor $\chi^{(3)}$ nonlinear interactions to realize these distinct functionalities within a common platform (Fig.~1). The quantum source uses spontaneous four-wave mixing (SFWM) in a stoichiometric silicon nitride (Si$_3$N$_4$) microring resonator to produce temporally-correlated photon pairs~\cite{caspani_integrated_2017}, while quantum frequency conversion is implemented through four-wave mixing Bragg scattering (FWM-BS)~\cite{mckinstrie_translation_2005} in a similar Si$_3$N$_4$ microring~\cite{li_efficient_2016,singh_quantum_2019}. The successful combination of photon pair generation and frequency conversion in a common platform enables frequency conversion of a quantum state of light and establishes its relevance to frequency-bin-encoded quantum photonics~\cite{olislager_frequency-bin_2010,lukens_frequency-encoded_2017,kues_-chip_2017,imany_50-ghz-spaced_2018}. Here, the inherent high-dimensionality in which photon pairs are distributed amongst frequency modes necessitates methods to implement controllable frequency shifts for efficient mixing of frequency bins. To that end, we use QFC to remove the spectral distinguishability between the two nondegenerate photons comprising a photon pair, and demonstrate tunable Hong-Ou-Mandel interference in which the quantum beat of single photons is observed~\cite{legero_quantum_2004}.

\textit{System Overview}\textemdash Figure 1 provides an overview in which one photon from a non-degenerate microresonator photon pair source is sent to a microresonator frequency converter. Measurements establish the preservation of quantum correlations and the ability to use frequency conversion to enable quantum interference. In contrast to typical approaches for quantum frequency conversion (QFC)~\cite{kumar_quantum_1990,raymer_manipulating_2012} that employ centimeter-scale nonlinear crystals or meter-scale optical fibers~\cite{tanzilli_photonic_2005,rakher_quantum_2010,mcguinness_quantum_2010,zaske_visible--telecom_2012,ates_two-photon_2012,albrecht_waveguide_2014,clemmen_ramsey_2016,wright_spectral_2017,walker_long-distance_2018,maring_quantum_2018,dreau_quantum_2018,siverns_neutral_2018}, rendering compact integration with quantum nodes difficult, our implementation of both the quantum source and frequency converter uses chip-integrated, low-power silicon nanophotonics.

Achieving efficient $\chi^{(3)}$ nonlinear processes in microresonators~\cite{moss_new_2013} requires phase- and frequency-matching of the four involved fields and efficient coupling to the resonators at the signal, idler, and pump wavelengths. To integrate nanophotonic elements based on different nonlinear processes, these criteria must be satisfied for each element while keeping device geometries compatible. The Si$_3$N$_4$ microring we optimize for FWM-BS shows small normal dispersion in both the 930~nm and 1550~nm bands (Section I of the Supplementary Material (SM)). As a result, photon pairs can be generated in microrings of similar dimensions (most critically, the same device layer thickness), by simply pumping with a relatively strong power (mW-level) in either of these two near-zero dispersion bands, with SFWM resulting in annihilation of two pump photons and creation of quantum-correlated signal/idler pairs. Because the photon pair source and frequency converter are implemented with the same device layer thickness, they can in principle be integrated on the same chip, ideally with filters, phase shifters, detectors, and other photonic elements that have been demonstrated in Si$_3$N$_4$~\cite{barwicz_microring-resonator-based_2004,xiong_compact_2015,schuck_nbtin_2013,xue_thermal_2016}. To retain flexibility in independent characterization of the source and frequency converter, here they are on separate chips, but there is no significant barrier to future direct integration.

\begin{center}
	\begin{figure*}
		\begin{center}
			\includegraphics[width=0.75\linewidth]{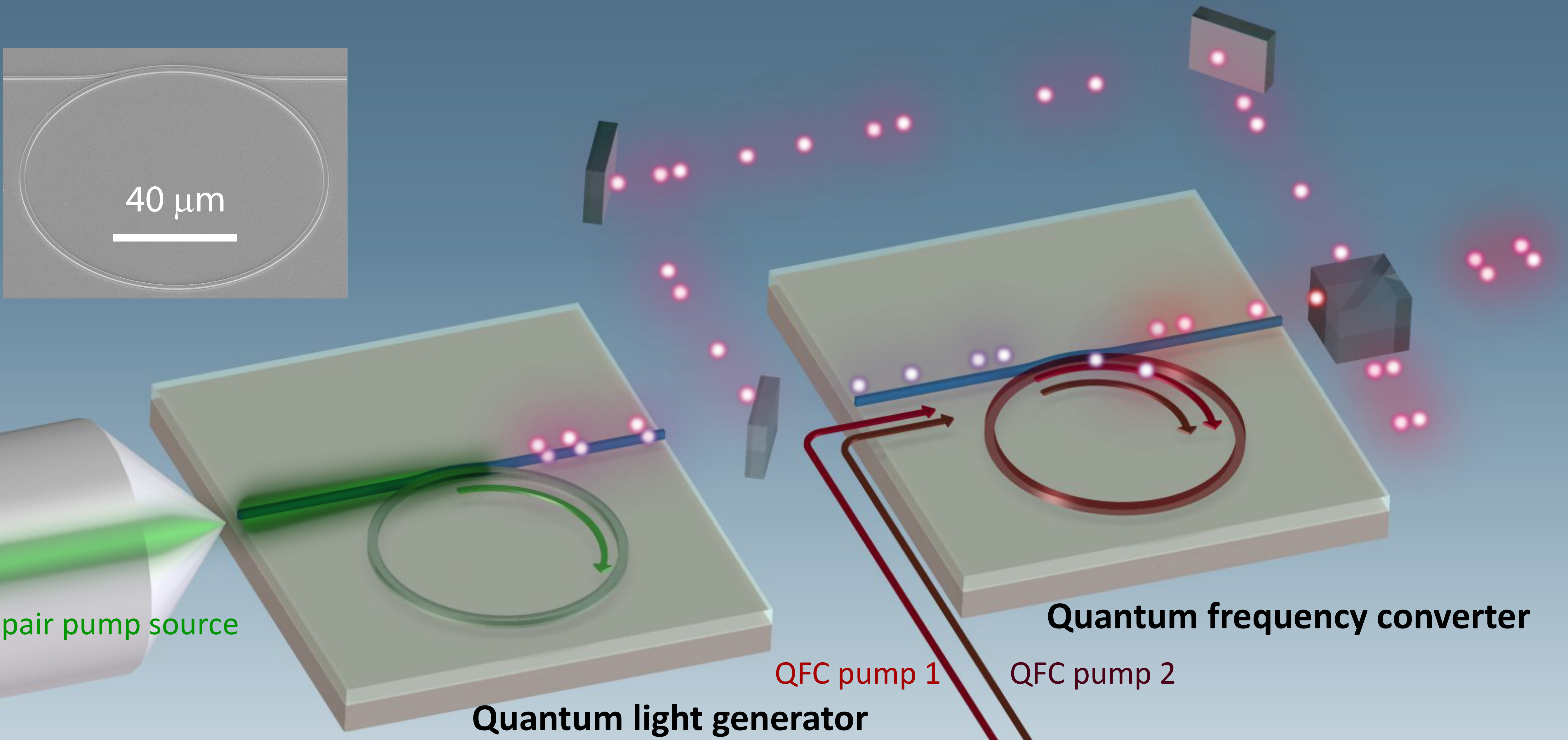}
			\caption{\textbf{Quantum beats using nonlinear nanophotonics}. Photon pairs are generated by spontaneous four-wave mixing in a Si$_3$N$_4$ microring. Signal photons are sent to the frequency converter, where they are spectrally translated to the vicinity of the idler frequency through four-wave mixing Bragg scattering in a second Si$_3$N$_4$ microring. A tunable quantum beat is observed in the Hong-Ou-Mandel interference of the idler and the frequency converted signal, with a signature that depends on their frequency difference.}
			\label{Figure1}
		\end{center}
	\end{figure*}
\end{center}

\textit {Devices}\textemdash The photon pair source is implemented in 40~$\mu$m radius, 500~nm thick Si$_3$N$_4$ microrings with ring widths around 1500 nm (Fig.~1). Due to the aforementioned small dispersion in the 930 nm band, we expect to observe a quantum comb consisting of a number of signal and idler frequency bins, as illustrated in Fig.~2a and experimentally observed in Fig.~2b. While such a quantum comb can exhibit high-dimensional entanglement across these frequency bins~\cite{kues_-chip_2017}, here we spectrally filter to work with one signal-idler combination at a given time (highlighted in Fig.~2b). We perform source characterization by passively transmitting the signal photons through the subsequent frequency converter chip (i.e., without applying the pumps that enable frequency conversion) and then onto a single-photon detector, while idler photons are directly sent to a second single-photon detector, with the photon flux and signal-idler coincidences recorded (see Supplementary Material Section IV for the experimental setup). The detected signal photon flux (Fig.~2b) is~$\approx~3.6\times 10^5$ s$^{-1}$ for a pump power $\approx$~4~mW, while the on-chip pair generation rate is~$\approx~2\times 10^7$~s$^{-1}$ after accounting for losses. We further characterize the source by measuring its linewidth ($\approx$~640~MHz) and coincidence-to-accidental ratio (CAR), defined as the ratio between the peak coincidence contrast and noise background, as a function of pump power. The CAR corresponding to Fig.~2b is $\approx$~10 (details provided in SM Section II).

\begin{center}
	\begin{figure}
		\begin{center}
			\includegraphics[width=\linewidth]{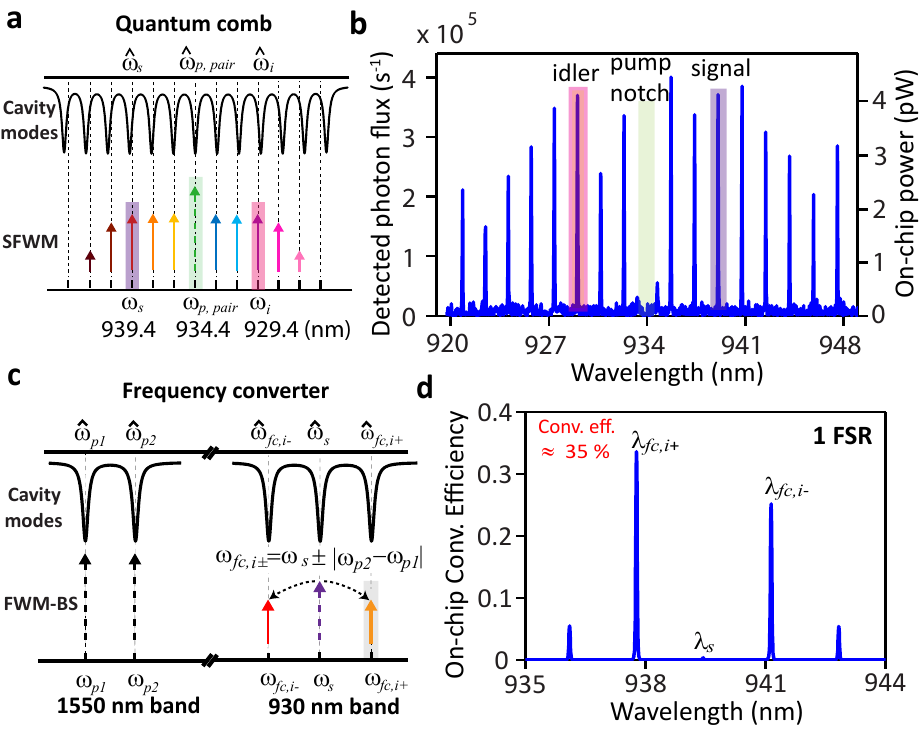}
			\caption{\textbf{Microresonator photon pair source and frequency converter}. \textbf{a}, Schematic of the SFWM process in a microring, where quantum-correlated signal and idler photons are distributed amongst multiple frequency bins. \textbf{b}, Measured pair spectrum for an input power near 4 mW, where the left $y$ axis is the detected signal photon flux and the right $y$ axis is the corresponding on-chip power. The highlighted  pair in \textbf{a}-\textbf{b} is chosen as the signal and idler for the quantum frequency conversion experiment in Fig.~3. \textbf{c}, Intraband conversion scheme based on FWM-BS, where the signal is translated to nearby resonances with the frequency shift determined by the frequency difference of two pump lasers in the 1550 nm band. \textbf{d}, Output spectrum of the frequency converter with a 1 FSR separation between the two pump lasers (frequency translation $\approx 572$ GHz) for a total pump power of 20 mW on chip. The power in the 930 nm band is normalized by the input signal power. The input signal is a narrow linewidth, continuous-wave laser.}
			\label{Figure2}
		\end{center}
	\end{figure}
\end{center}
\vspace{-0.1in}

The frequency converter is based on FWM-BS in a 40 $\mu$m radius, 500~nm thick Si$_3$N$_4$ microring with 1450~nm ring width. In FWM-BS, the spectral shift is set by the difference in frequencies of two applied pumps~\cite{mckinstrie_translation_2005}. We focus on intraband translation due to its relevance to quantum comb sources. Here, the two pumps are on resonance with cavity modes in the 1550 nm band (cavity free spectral range, or FSR, is 0.52~THz), with input signal photons in the 930~nm band (Fig.~\ref{Figure2}c). By optimizing the device performance, we observe that for an input signal with bandwidth much smaller than the cavity linewidth (in this case, a continuous-wave (cw) laser $<$200~kHz linewidth), the signal can be completely depleted (Fig.~2d). The first-order blue-shifted idler ($\textit{fc,i}+$) has an on-chip conversion efficiency around $35~\%$, whereas the conversion efficiency of the first-order red-shifted idler ($\textit{fc,i}-$) is around $25~\%$. This asymmetry arises from the slight frequency mismatch between the amount of frequency shift (determined by the pump separation) and the FSR in the 930 nm band, and can be used to enhance the power of one particular idler. Second-order idlers are also observed, but with significantly reduced conversion efficiency. A systematic investigation of conversion efficiency, bandwidth, idler asymmetry and on-chip noise as a function of the pump power is presented in SM Sec. III.

\textit {Quantum frequency conversion}\textemdash We next consider the combined operation of the quantum source and frequency converter. The frequency converter is operated at a total applied pump power of 20~mW, corresponding to a 2 GHz converter bandwidth and 3~fW on-chip noise. This provides adequate conversion bandwidth (compared to the pair source bandwidth of 640~MHz) while maintaining sufficiently low noise. Adjusting the frequency converter temperature enables spectral matching of its relevant mode to the pair source signal photon (at $\lambda_{s}~\approx~939$~nm). Figure 3a shows the output spectrum corresponding to 1 FSR frequency translation, where the remnant photon pair source signal is several times smaller than the two frequency-converted idlers. Its suppression is weaker than the cw case (Fig.~2d) due to its broader linewidth, which is an appreciable fraction of the 2~GHz conversion bandwidth. For the same reason, the conversion efficiency of the blue idler has degraded from $35~\%$ in the cw case to $25~\%$, as expected given the input signal bandwidth (detailed analysis in SM Sec. III).

\begin{center}
	\begin{figure}[t]
		\begin{center}
			\includegraphics[width=\linewidth]{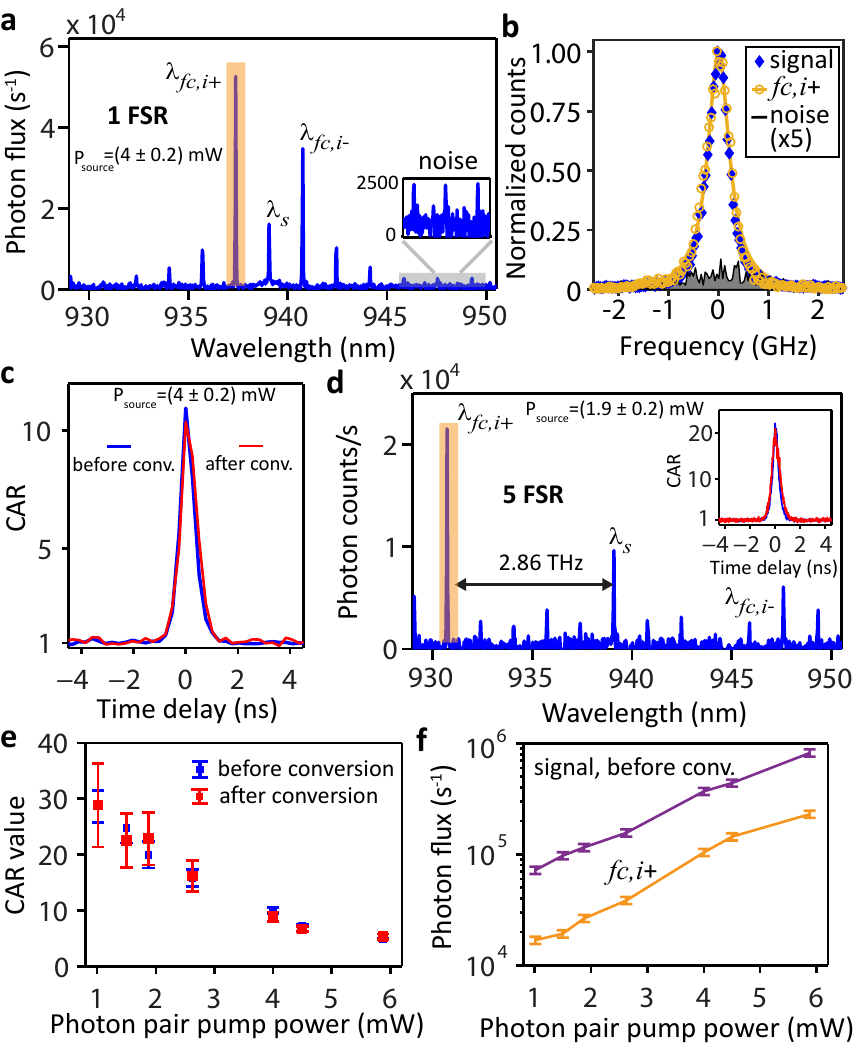}
			\caption{\textbf{Quantum frequency conversion}. \textbf{a}, Frequency-converted spectrum in which signal photons from the photon pair source (see Fig.~2b) are spectrally translated by 1 FSR. The inset shows a zoom-in plot of the resonant noise induced by the two 1550 nm pumps. \textbf{b}, High-resolution spectra of the input signal, frequency-converted blue idler highlighted in \textbf{a}, and resonant noise. The resonant noise is detected by removing the input signal. It shares a common normalization with the frequency-converted blue idler, and is enlarged 5 times in the plot for clarity. \textbf{c}, Measured CAR values corresponding to \textbf{a}. The blue curve is taken for the pair source signal and idler before conversion, while the red curve is for the frequency-converted blue idler and original pair source idler. \textbf{d}, Spectrum for 5 FSR frequency separation and a reduced photon pair source flux (pumped at 1.9~mW instead of 4~mW). The inset shows the corresponding CAR values before (blue) and after (red) conversion. \textbf{e}, Comparisons of CAR values before and after conversion as a function of photon pair source pump power. \textbf{f}, Detected flux of the pair source signal before conversion and the frequency-converted blue idler after conversion (1~FSR shift as in \textbf{a}). The on-chip conversion efficiency is estimated to be $25~\% \pm 3~\%$. The converted photon flux has been accounts for $50~\%$ loss from the bandpass filter used to select the blue idler.}
			\label{Figure3}
		\end{center}
	\end{figure}
\end{center}

\begin{center}
	\begin{figure*}
		\begin{center}
			\includegraphics[width=0.75\linewidth]{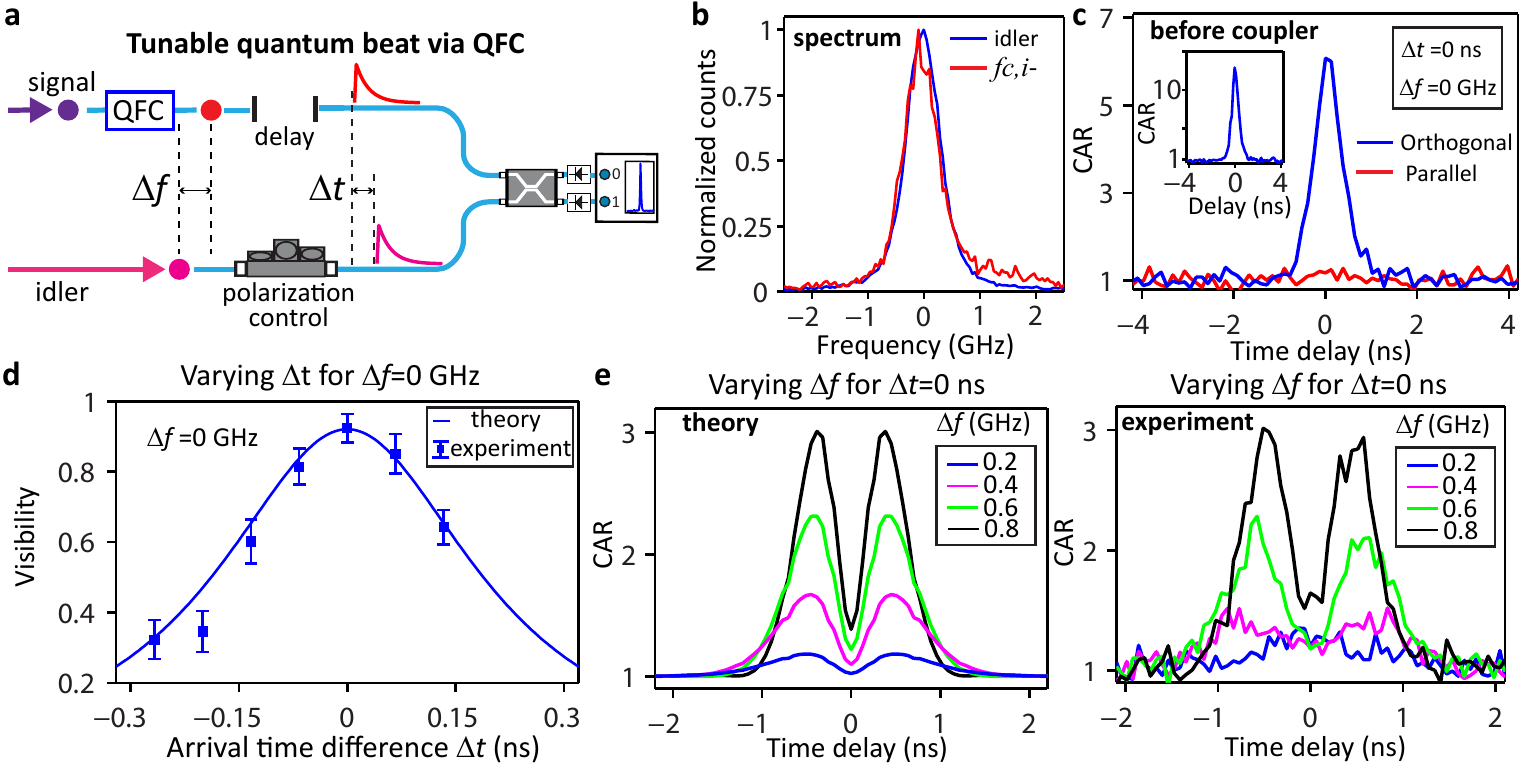}
			\caption{\textbf{Tunable quantum beat enabled by QFC}. \textbf{a}, Experimental schematic, where the frequency difference of the two photons ($\Delta f$), the relative arrival time on the coupler ($\Delta t$), and the relative polarization can all be varied. \textbf{b}, Superimposed high-resolution spectra for the pair source idler and the frequency-converted idler. \textbf{c}, Coincidence counting after the 50/50 coupler, taken for $\Delta t$=0 and $\Delta f$=0 , for both orthogonal (blue) and parallel (red) polarizations. The inset shows the coincidence counting measurement for the two photons before entering the 50/50 coupler. \textbf{d}, Theoretical (solid curve) and measured (markers with error bars) visibility as a function of the relative optical delay between two photons, with $\Delta f$=0. \textbf{e}, Theoretical (left) and experimental (right) coincidence counting corresponding to $\Delta t$=0 and varied frequency difference. Only the co-polarized case is considered in \textbf{d-e}.}
			\label{Figure4}
		\end{center}
	\end{figure*}
\end{center}

Figure 3b shows a high-resolution spectrum of the frequency-converted blue idler (highlighted in Fig.~3a), displaying essentially the same bandwidth as the original pair source signal. In addition, a weak resonance-shaped noise peak at the same frequency is detected by blocking the input signal but continuing to apply the frequency converter pumps. This noise is also observable in the inset to Fig.~3a. In Fig.~3c, the CAR after conversion (red curve), which correlates the frequency-converted idler and the original pair source idler, is compared against the CAR before conversion (blue curve). The almost identical responses suggest that the additional noise introduced in the QFC process is small relative to the signal level. In Fig.~3d, we increase the frequency separation between the two pumps to 5 FSRs, and a similar conversion efficiency near $25~\%$ is obtained for the corresponding 5-FSR-shifted blue idler. Compared to Fig.~3a, the pump power for the photon pair source has been reduced, resulting in a smaller photon flux but an increased CAR near 20 both before and after conversion (inset in Fig.~3d). Finally, a scan of the pump power for the photon pair source is performed while keeping the configuration of the frequency converter fixed. The results (Fig.~3e) demonstrate that the quantum correlation between the photons in the pair source is preserved by our frequency converter (within the measurement error), with a uniform conversion efficiency $25~\% \pm 3~\%$ (Fig.~3f) for the full range of input photon flux. Additional analysis of noise in the QFC process is presented in SM Sec. V.

\textit {Quantum Beat of Single Photons} \textemdash Having demonstrated QFC, here we use it to observe the tunable quantum beat of single photons. This is accomplished by using a photon pair source whose FSR is nearly the same as that of the frequency converter. The experimental scheme is illustrated in Fig.~4a, where the photon pair source signal and idler are initially separated by 2 FSRs ($\approx~1.14$~THz). The signal photons are then brought spectrally close to idler photons through QFC. While the frequency converter FSR largely determines this spectral shift, fine tuning can be achieved because the cavity modes have a finite linewidth. That is, we can tune the pump lasers within their respective cavity mode linewidths (few hundred MHz each) to achieve the precise spectral shift needed for high-visibility interference (SM Sec VI).

Figure 4b confirms that for an optimized frequency shift, the original pair source idler and the frequency-converted red idler spectrally overlap (i.e., $\Delta f=0$). Since these two photons are nearly identical and strongly correlated in time, we can send them to a 50/50 fiber coupler and perform a two-photon interference experiment. First, we ensure that two photons arrive at the coupler at the same time through a tunable optical delay line (Fig.~4a). By adjusting the polarization of the pair source idler while fixing that of the frequency-converted idler, two distinct responses in the coincidence measurement are observed. If the two photons are orthogonally polarized (blue curve in Fig.~4c), we observe a coincidence peak whose CAR value is half what it was before entering the 50/50 coupler (inset to Fig.~4c). On the other hand, if we adjust their polarization to be the same, the coincidence peak disappears to the background accidental coincidence level (red curve in Fig.~4c), indicating destructive interference. Essentially, there are two possibilities corresponding to the detection of one photon at each detector: each photon exits from its through port of the coupler, or each photon exits from its cross port. The probability amplitudes associated with these two scenarios cancel, resulting in an accidentals-level coincidence rate, i.e., the Hong-Ou-Mandel effect~\cite{hong_measurement_1987}.

Next, we vary the frequency difference ($\Delta f$) and the relative optical delay ($\Delta t$) between the pair source idler and the frequency-converted idler. In the first case, the two photons share the same frequency but arrive at the 50/50 coupler at different times (i.e., $\Delta t\neq0$ and $\Delta f=0$). To characterize this process, we define a visibility based on the contrast of peak values in the coincidence rate between the parallel and orthogonal polarizations. The experimental data (Fig.~4d), showing a maximum visibility $>90~\%$, agrees with the simulation result reasonably well (see SM Sec. VII for theoretical modeling). In the second case, the two photons arrive at the coupler at the same time but with different frequencies (i.e., $\Delta t=0$ and $\Delta f \neq 0$). Here, the normalized coincidence rate (by the orthogonal polarization case) for the parallel polarization is given by $1-\cos(2\pi\Delta f \tau)$, where $\tau$ is the relative electronic delay between the two detectors (i.e., the $x$ axis in Fig.~4e). This formula indicates that even if the two photons have different frequencies, there is an interference dip at $\tau=0$, which has been confirmed by experimental results shown in Fig.~4e. As can be seen, the interference dip fails to reach the accidental coincidence level for large $\Delta f$, which is mainly due to the limited detector timing resolution (120 ps) and time bin (64 ps) used in the coincidence counting. After taking these factors into consideration, a reasonable agreement between theory and experiment is observed. Oscillations in the two-photon interference of single photons with slightly different frequencies has been referred to as a quantum beat~\cite{legero_quantum_2004}, and shows the sensitivity of two-photon interference to precise spectral matching~\cite{legero_characterization_2006}. In comparison to those previous works, here we control the quantum beat through QFC.

\textit {Summary} \textemdash We have thus demonstrated the generation and frequency conversion of quantum states of light in a nonlinear nanophotonic platform. The high visibility and controllable frequency differences achieved indicate that nanophotonic QFC could enable the high-quality quantum interference needed for various applications. Going forward, the frequency-bin-entangled states produced by microresonator SFWM and their synergy with QFC makes the platform appealing for exploring sophisticated frequency domain manipulation of quantum states. Operation of the FWM-BS device as a frequency tritter~\cite{lu_electro-optic_2018} and as a nonlinear element addressing all frequency bins simultaneously are amongst the opportunities enabled by our platform.

\begin{acknowledgments}
Q.~Li, A.~Singh, X.~Lu acknowledge support under the Cooperative Research Agreement between the
University of Maryland and NIST-CNST, Award 70NANB10H193. The authors wish to thank Marcelo Davan\c{c}o from NIST Gaithersburg and Abijith Kowligy from NIST Boulder for helpful comments.
\end{acknowledgments}

\onecolumngrid \bigskip
\setcounter{figure}{0}
\setcounter{equation}{0}
\makeatletter
\renewcommand{\theequation}{S\@arabic\c@equation}
\begin{center} {{\bf \large SUPPLEMENTARY
MATERIAL}}\end{center}

\renewcommand{\thefigure}{S\arabic{figure}}

\section{Linear transmission and dispersion characterization}

\vspace{-0.25in}

\begin{center}
	\begin{figure*}[b!]
		\begin{center}
			\includegraphics[width=0.8\linewidth]{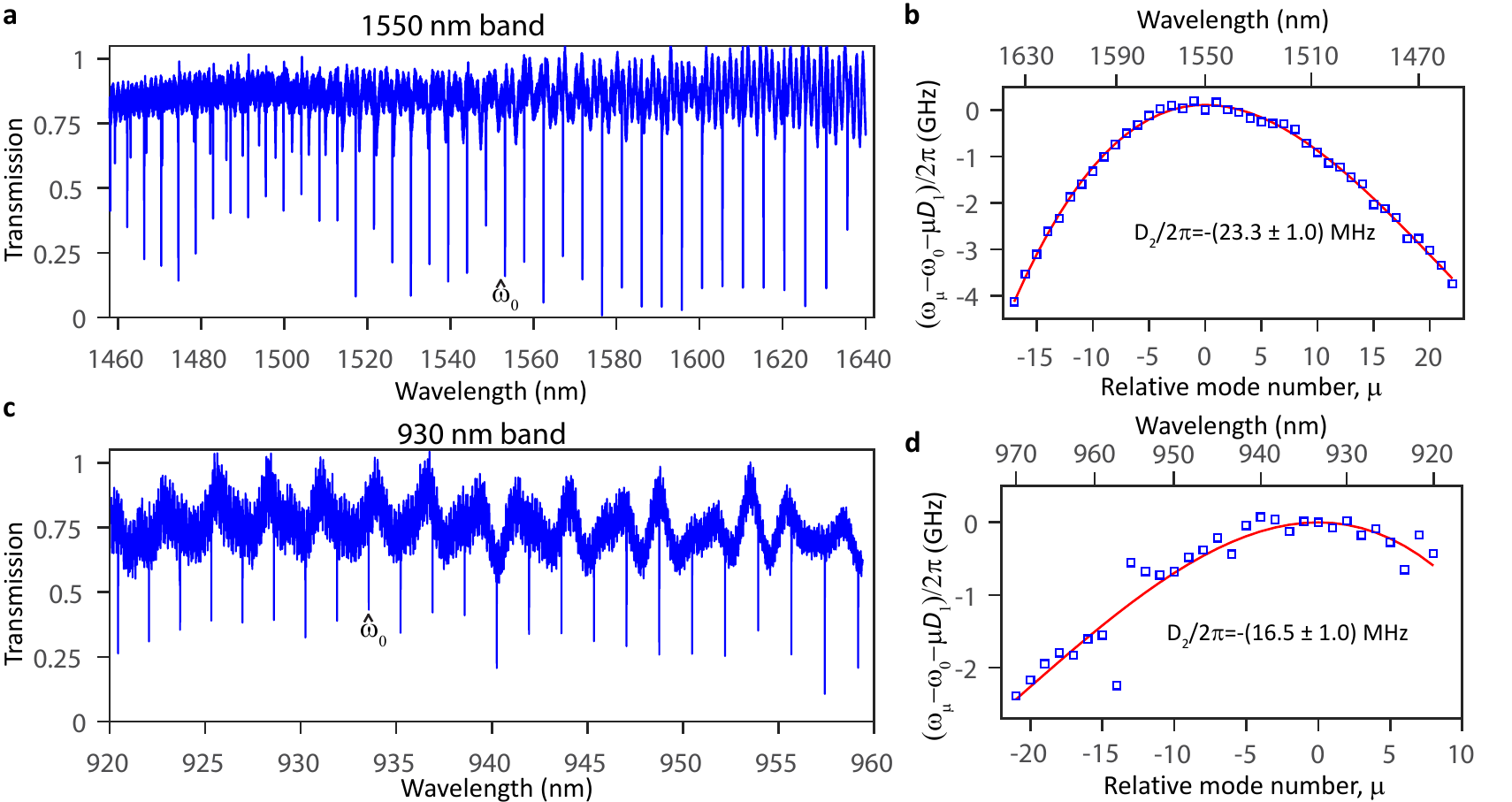}
			\caption{\textbf{Linear transmission and dispersion of a representative Si$_3$N$_4$ microring}. \textbf{a}, Linear transmission scan of a representative device (used as the frequency converter in this work) in the 1550 nm band, where $\hat{\omega}_0$ marks the reference frequency for dispersion determination. \textbf{b}, Deviation of the measured resonance frequencies (markers) from an equidistant frequency grid $\hat{\omega}_0 + \mu D_1$ in the 1550 nm band. The extracted $D_2$ is provided, where the displayed one standard deviation uncertainty is determined from a nonlinear least squares fit to the data. \textbf{c}, Linear transmission scan of the same device as \textbf{a} in the 930 nm band, where the resonances are over-coupled to achieve high conversion efficiencies. \textbf{d}, Deviation of the measured resonance frequencies (markers) from an equidistant frequency grid $\hat{\omega}_0 + \mu D_1$ in the 930 nm band. Coupling between the mode of interest and another mode of the resonator at $\mu\approx-13$ causes a deviation from quadratic behavior; these points are not included in the fit.}
			\label{EDF_Linear}
		\end{center}
	\end{figure*}
\end{center}

The 40 $\mu$m radius Si$_3$N$_4$ microring resonators employed in this work, both for the photon pair generation and frequency conversion, are similar to the ones adopted in our earlier work~\cite{li_efficient_2016}. The Si$_3$N$_4$ thickness is around 500 nm, and the ring width has been varied from 1400 nm to 1600 nm with a step size of 10 nm. The coupling between the resonator and access waveguide is based on a pulley coupling scheme, which can be engineered by varying the access waveguide width, gap, and pulley coupling length. Optimization of these parameters allows us to tailor the properties of microring resonators, in terms of their FSR, dispersion, and coupling, for each specific application. For example, there are two essential requirements for intraband frequency conversion in the 930 nm band: (1) identical FSRs between the 1550 nm and 930 nm bands to satisfy frequency matching (for modes that are phase-matched), and (2) overcoupling of the resonances in the 930 nm band to attain overall high conversion efficiencies. In addition, the resonances in the 1550 nm are preferred to be critically coupled so the required pump powers are minimized.
	
The transmission of a typical microring resonator, both in the 930 nm and 1550 nm bands, is provided in Fig.~\ref{EDF_Linear}. The resonance frequencies in each band can be approximated by the Taylor series $\hat{\omega}_\mu=\hat{\omega}_0 + D_1\mu+ \frac{1}{2}D_2\mu^2+\dots$, where $\hat{\omega}_0$ is the reference frequency, $D_1/2\pi$ is the FSR of the resonator at $\hat{\omega}_0$, $D_2$ is the quadratic dispersion, and $\mu$ is an integer representing the relative mode order number with respect to $\hat{\omega}_0$. As shown in Fig.~\ref{EDF_Linear}, the dispersion in both bands is found to be normal ($D_{2}<0$) and small enough to enable photon pair generation in either band by simply pumping the resonance at a relatively high power. While the resonances in the 930 nm band of this representative device are overcoupled to achieve high conversion efficiencies for the frequency conversion process, for the photon pair generation we use a slightly modified waveguide-resonator coupling design, choosing it to be near critical coupling in the 930 nm band. This ensures that the spectral bandwidth of the emitted photon pairs is several times smaller than the frequency conversion bandwidth.

\vspace{0.1in}
\section{Photon pair source characterization}
\begin{center}
	\begin{figure*}[h]
			\begin{center}
				 \includegraphics[width=0.9\linewidth]{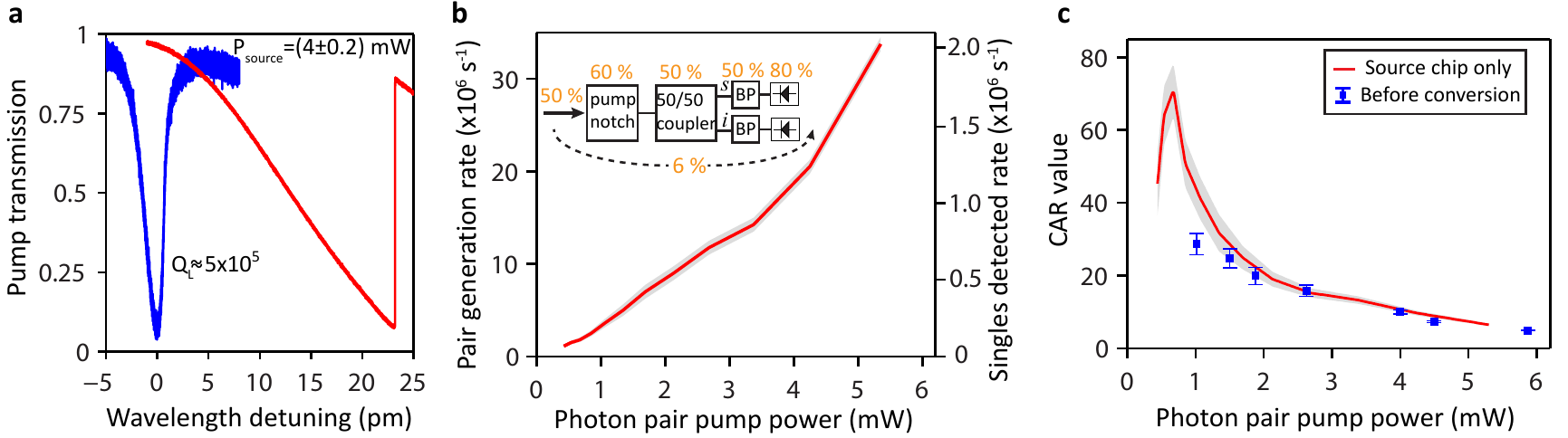}
				 \caption{\textbf{Photon pair source characterization}. \textbf{a}, Transmission of the pump resonance in the linear case (blue curve) and at a pump power near 4 mW (red curve). \textbf{b}, Detected signal (or idler) photon flux of the third-order pair (right y-axis) and the inferred on-chip pair generation rate (left y-axis) after accounting for transmission losses. The red solid line denotes the mean value of the photon flux and the gray area denotes the standard deviation extracted from the measured counts. The inset shows the experimental schematic and the transmission of relevant components is provided, including: waveguide out-coupling ($50~\%$), pump notch filter ($60~\%$), 50/50 coupler to split photons to two channels for coincidence measurement ($50~\%$), bandpass filters to select the desired signal and idler ($50~\%$), and the superconducting single photon detectors ($80~\%$ detection efficiency). Overall, approximately $6~\%$ of the photons emitted from the on-chip waveguide are detected. \textbf{c}, The red solid line denotes the measured coincidence-to-accidental ratio (CAR) for the photon pair source chip based on the experimental schematic shown in \textbf{b}, with the gray area denoting the standard deviation extracted from the measured counts. We have also plotted the measured CAR values when the signal is passively transmitted through the frequency converter (i.e., no frequency conversion), corresponding to ``before conversion" case shown in Fig.~3e in the main text.}
 \label{FigS_photon_source}
 \end{center}
 \end{figure*}
 \end{center}

As shown in Fig.~\ref{FigS_photon_source}a, the resonances for the photon pair generation microring in the 930 nm band are critically coupled with a loaded quality factor $Q\approx 5\times 10^5$ in the linear regime. This corresponds to a spectral bandwidth of~$\approx$~640 MHz, consistent with the result from the scanning Fabry-Perot measurement (see Fig.~3b in the main text). With increased pump powers, the resonator starts to generate an appreciable photon pair flux while eventually exhibiting thermal bistability in the pump transmission. In Fig.~\ref{FigS_photon_source}b, we plot the detected signal (or idler) photon flux (right $y$ axis) as a function of the pump power, from which the on-chip pair generation rate is estimated after accounting for various losses in the transmission link (see the inset shown in Fig.~\ref{FigS_photon_source}b). Subsequently, the coincidence-to-accidental ratio (CAR) for the selected signal and idler pair is measured (red solid line in Fig.~\ref{FigS_photon_source}c), which has a peak value around 70 at a pump power near 0.67 mW. We have also plotted the CAR measurement by transmitting the signal through the frequency conversion chip without applying any pumps (blue markers in Fig.~\ref{FigS_photon_source}c, corresponding to the ``before conversion" case shown in Fig.~3e in the main text). As can be seen, the additional loss in the signal and idler photons incurred from passing through the frequency converter does not degrade the CAR value much for the range of pump power tested. A more detailed study analyzing the impact of various noise sources (including the Raman noise) on the CAR measurement is provided later in Section~V.

\section{Frequency converter: additional discussion}

\subsection{Characterization}

\begin{center}
	\begin{figure*}[b]
		\begin{center}
			\includegraphics[width=0.9\linewidth]{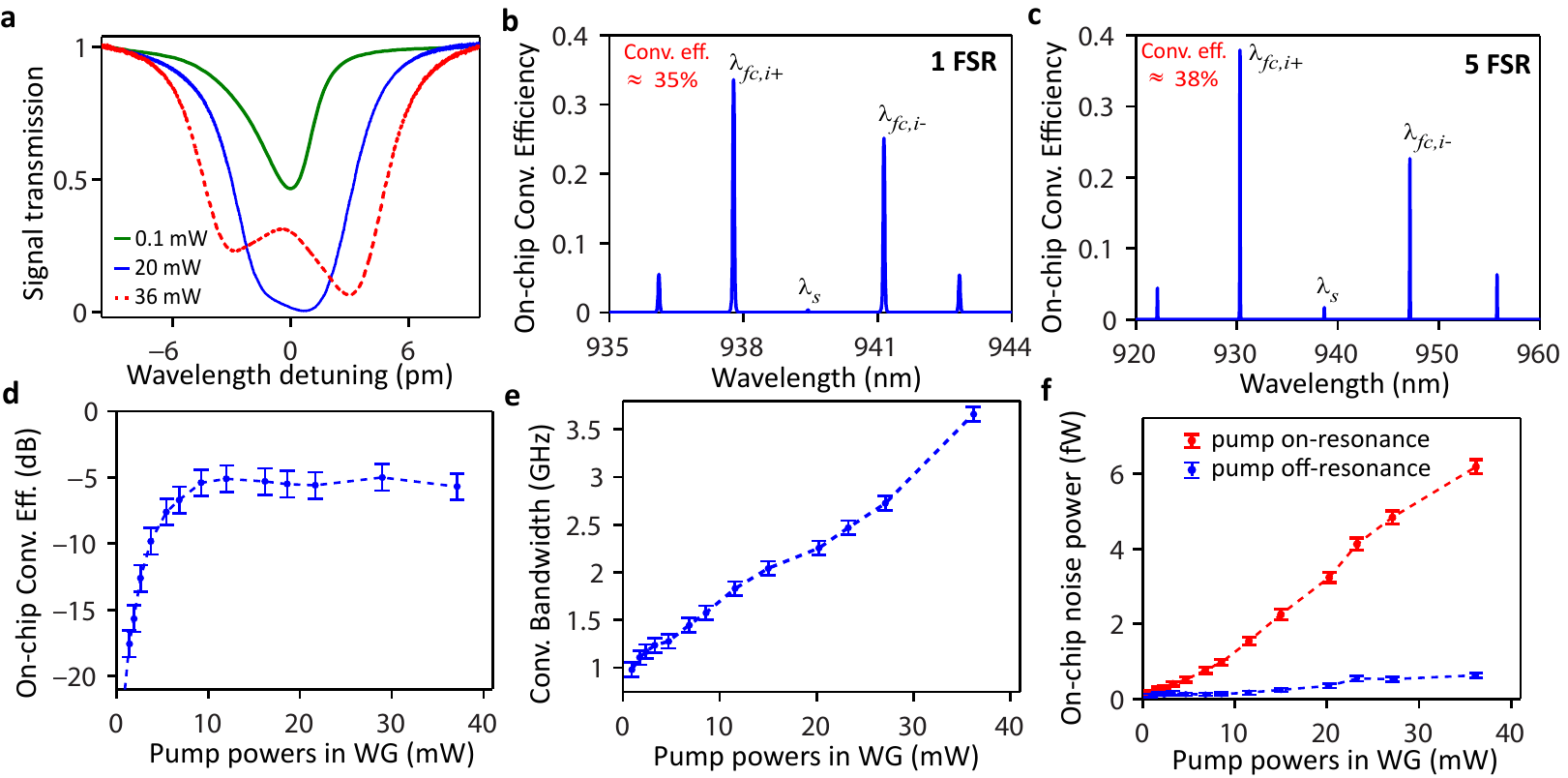}
			\caption{\textbf{Characterization of the microresonator frequency converter.}. \textbf{a}, Signal transmission based on a swept tunable laser for several different pump powers. \textbf{b}, Output spectrum of the frequency converter with a 1 FSR separation between the two pump lasers (frequency translation $\approx$~572 GHz) for a total pump power of 20~mW on chip (10~mW each). The power in the 930 nm band is normalized by the input signal power, corresponding to the on-chip conversion efficiency. \textbf{c}, Frequency conversion spectrum for 5 FSR separation between two pump lasers (frequency translation $\approx$~2.86 THz) for a total pump power of 20~mW on chip (10~mW each). The next three figures show dependence on total on-chip pump power of the \textbf{d}, On-chip conversion efficiency of the blue-shifted idler fc,i+ , \textbf{e}, Conversion bandwidth, and \textbf{f}, On-chip noise. Error bars in \textbf{d}-\textbf{e} represent one standard deviation values caused by fluctuations in the detected signal. }
			\label{EDF_FC}
		\end{center}
	\end{figure*}
\end{center}

The frequency converter devices are initially characterized using a narrow linewidth, continuous-wave laser for the input signal.  Figure~\ref{EDF_FC}a shows the signal resonance for three different pump powers, going from the overcoupled regime at low pump powers to the critically-coupled and undercoupled regimes at increased pump powers, as the input signal is depleted due to frequency conversion into output idlers. The width of the signal resonance also increases with increasing pump power, indicating that the conversion bandwidth can be broadened beyond the microresonator's loaded linewidth in the linear regime.  The spectra of 1 FSR separation and 5 FSR separation are provided in Fig.~\ref{EDF_FC}b and Fig.~\ref{EDF_FC}c, respectively, showing a consistent conversion efficiency above 35~$\%$ for the blue idler.

The pump power dependence of the conversion efficiency, bandwidth, and generated noise power are shown in Fig.~\ref{EDF_FC}d, ~\ref{EDF_FC}e, and ~\ref{EDF_FC}f, respectively.  As can be seen in Fig.~\ref{EDF_FC}d, the conversion efficiency saturates at $\approx~30~\%~\pm5~\%$ for total pump powers above 10~mW. However, both the conversion bandwidth and the noise associated with the FWM-BS process also increase with the pump power. This noise is believed to stem from fluorescence centers in the Si$_3$N$_4$ material. It is much stronger when the pumps are aligned with their respective cavity modes, a consequence of the resonance-enhancement of the pump intensities. The optimal choice of pump power depends on the experiment, where the input signal bandwidth, input signal flux, and desired signal-to-noise level for the frequency-converted idler are taken into consideration.
\vspace{-0.15in}

\subsection{Conversion efficiency vs. signal input bandwidth}

The photon pair source used in the quantum frequency conversion experiment differs from the cw laser used in the classical characterization in that the pair photons have a much larger bandwidth (640~MHz vs. 200~kHz). In this subsection, we investigate the impact of such a finite bandwidth on the conversion efficiency by employing a simplified set of coupled mode equations developed in our earlier work~\cite{li_efficient_2016}:
\begin{align}
t_R\frac{d E_{s}}{dt} &= -(\alpha + i\delta_s )E_s + i\Omega_0 E_{i-} + i \Omega_0 E_{i+} + i \sqrt{\theta P_s}, \label{Eq_CMT_s} \\
t_R\frac{d E_{i+}}{dt} &= -\left(\alpha + i(\delta_s + \Omega_1+\Omega_2)\right)E_{i+} + i \Omega_0 E_s, \label{Eq_CMT_i1} \\
t_R\frac{d E_{i-}}{dt} &=  -\left(\alpha + i (\delta_s - \Omega_1 +\Omega_2)\right)E_{i-} + i \Omega_0 E_s,
\label{Eq_CMT_i2}
\end{align}
where $E_{s,i\pm}$ are the intracavity mean fields corresponding to the signal and two adjacent idlers ($|E|^2$ representing the average power traveling inside the cavity), $t_R$ is the round-trip time, $\alpha$ is the cavity loss rate in the 930 nm band ($\alpha=\hat{\omega}_st_R/(2Q_L)$ with $\hat{\omega}_s$ and $Q_L$ being the signal resonance frequency and its loaded $Q$, respectively), $\delta_s$ denotes the signal detuning, $\theta$ is the power coupling coefficient between the resonator and the access waveguide ($\theta=\hat{\omega}_st_R/Q_c$ with $Q_c$ being the coupling $Q$), and $P_s$ represents the power of a cw signal. The parameters $\Omega_n (n=0,1,2)$ are defined as:
\begin{align}
\Omega_0 & \equiv 2\gamma_{s}L|E_{p1}E_{p2}|, \label{Eq_Omega0}\\
\Omega_1 & \equiv \frac{\delta_{i+}-\delta_{i-}}{2},  \label{Eq_Omega1}\\
\Omega_2 & \equiv \frac{\delta_{i+}+\delta_{i-}-2\delta_s}{2}, \label{Eq_Omega2}
\end{align}
where $\gamma_s$ is the Kerr nonlinear coefficient in the 930 nm band, $L$ is circumference of the microring resonator ($L\equiv 2\pi R$ with $R$ being the ring radius), $E_{p1,p2}$ denote the intracavity mean fields of the two pumps in the 1550 nm band, and $\delta_{i\pm}$ are the detunings of the two idlers. A straightforward calculation shows that $\Omega_{1,2}$ can be expressed as:
\begin{gather}
\Omega_1  \approx  \left(D_1 |\mu| - |\omega_{p1}-\omega_{p2}|\right)t_R - \frac{\gamma_p L}{2} \left(|E_{p1}|^2 - |E_{p2}|^2\right),  \label{Eq_Omega1a}\\
\Omega_2  \approx \frac{1}{2}D_2 \mu^2 t_R,\label{Eq_Omega2a}
\end{gather}
where $\gamma_p$ is the Kerr nonlinear coefficient in the 1550 nm band.

\begin{center}
	\begin{figure*}[b]
		\begin{center}
			\includegraphics[width=\linewidth]{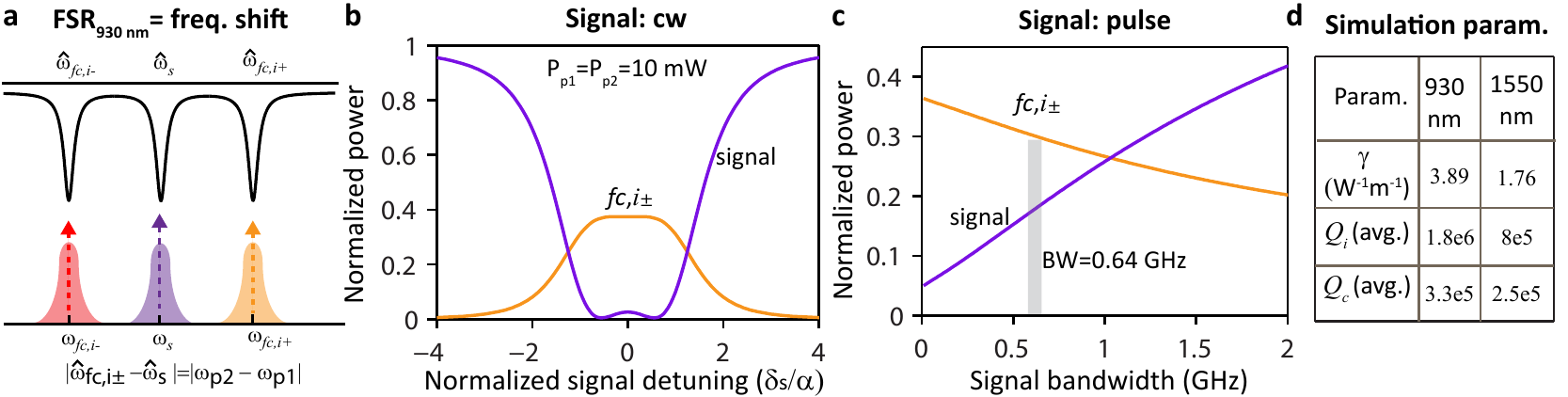}
			\caption{\textbf{Conversion efficiency vs signal bandwidth}. \textbf{a}, Schematic illustrating the condition for symmetric idlers: the FSR in the 930 nm band equals the frequency shift set by the two pump lasers in the 1550 nm band. \textbf{b}, Conversion efficiencies (orange) of the two idlers (equal in this case) and normalized transmission of the signal (purple) as a function of the signal detuning. \textbf{c}, Conversion efficiencies (orange) of the two idlers (equal in this case) and normalized transmission of the signal as a function of its input bandwidth at zero signal detuning. The gray bar highlights the 640~MHz bandwidth for the pair source used in this work. \textbf{d}, Table listing simulation parameters, including the nonlinear coefficient ($\gamma$), the intrinsic quality factor ($Q_i$) and the coupling quality factor ($Q_c$) in both the 930 nm and 1550 nm bands. The pump powers are assumed to be 20 mW in the 1550 nm band (10 mW for each pump).}
			\label{FigS_FC_BW}
		\end{center}
	\end{figure*}
\end{center}

It is obvious from Eqs.~\ref{Eq_CMT_s}-\ref{Eq_CMT_i2} that the two idlers are symmetric if $\Omega_1=0$. In our configuration, the two pumps usually have equal power. Thus, according to Eq.~\ref{Eq_Omega1a}, $\Omega_1$ is determined by the difference between the frequency shift, which is dictated by the frequency difference of the two pump lasers ($|\omega_{p1}-\omega_{p2}|$), and the FSR in the 930 nm band ($D_1$). For simplicity, here we limit our discussion to the 1 FSR case and consider the matched scenario first (Fig.~\ref{FigS_FC_BW}a). Figure ~\ref{FigS_FC_BW}b shows the simulated result for the cw case, where the signal is almost depleted (transmission less than $5~\%$) and each idler has a conversion efficiency around $38~\%$. Next, we extend the analysis to the finite bandwidth pulse by expanding its spectrum into a series of single frequency inputs (i.e., a Fourier expansion). As can be seen in Fig.~\ref{FigS_FC_BW}c, the conversion efficiency of the frequency converter gradually decreases with the increased bandwidth, whereas the signal transmission steadily increases. For example, for an input bandwidth of $0.64$ GHz which corresponds to the case of the photon pair source, the conversion efficiency degrades from $38~\%$ to $30~\%$, while the signal transmission increases from $5~\%$ to approximately $16~\%$. Further increasing the input bandwidth beyond 1 GHz will result in the signal being stronger than the two idlers, a clear indication that the signal cannot be sufficiently depleted.

\subsection{Asymmetric conversion efficiencies between two idlers}
\begin{center}
	\begin{figure*}[h]
		\begin{center}
			\includegraphics[width=\linewidth]{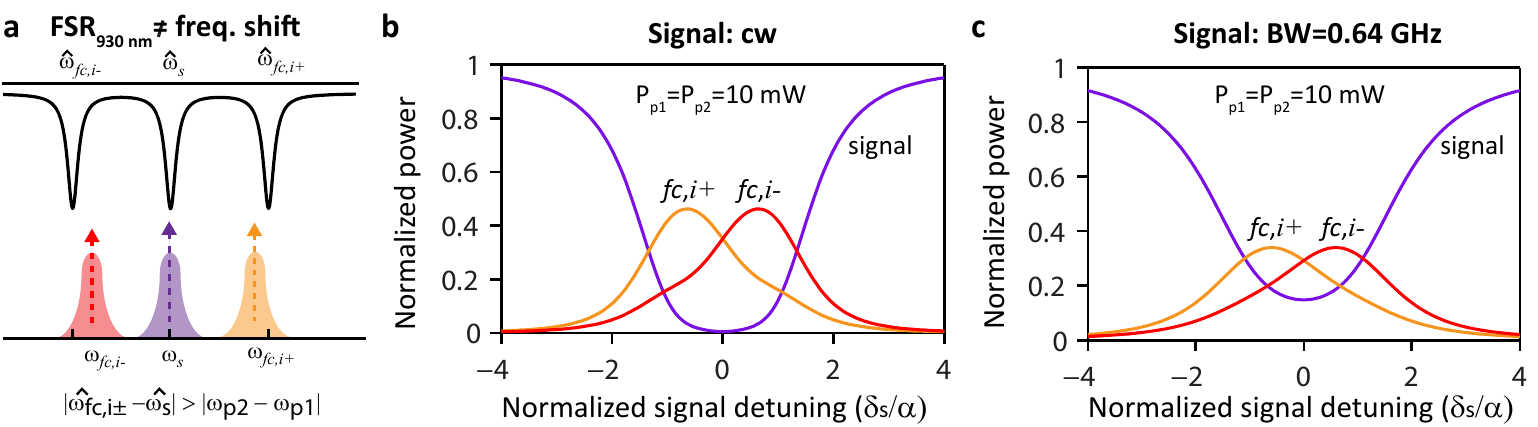}
			\caption{\textbf{Asymmetric conversion efficiencies in two idlers}. \textbf{a}, Schematic illustrating asymmetric idlers: the FSR in the 930 nm band is not equal to the amount of frequency translation set by the two 1550 nm pumps. \textbf{b}, Conversion efficiencies of the two idlers (orange and red) and normalized transmission of signal (purple) as a function of the signal detuning for $\Omega_1=0.5\alpha$, with other simulation parameters being the same as listed in Fig.~\ref{FigS_FC_BW}d. \textbf{c}, Conversion efficiencies of the two idlers (orange and red) and normalized transmission of the signal (purple) for an input bandwidth of $0.64$ GHz.}
			\label{FigS_FC_Detuning}
		\end{center}
	\end{figure*}
\end{center}

In this subsection, we consider the scenario that the frequency shift (set by the difference in pump frequencies) is not equal to the FSR in the 930 nm band.  We again limit our discussion to the 1 FSR case as illustrated in Fig.~\ref{FigS_FC_Detuning}a. In this case, the lack of symmetry in the frequency detunings of the two idlers (i.e., $\Omega_1\neq 0$) results in different conversion efficiencies for a given signal detuning. For example, Fig.~\ref{FigS_FC_Detuning}b plots the simulation results corresponding to the case of a cw signal, where the maximum achievable conversion efficiency for the blue or red idler is slightly higher (around $46~\%$) than the symmetric case ($38~\%$), due to the suppression of the other idler at that level of detuning. Similarly, for a pulsed input with a finite bandwidth (Fig.~\ref{FigS_FC_Detuning}c), we can adjust the relative strength between the blue and red idlers by varying the signal detuning, a feature that has been used in our experiment to increase the efficiency of the targeted idler.

We note that our use of a simplified coupled mode theory results in overestimates of the conversion efficiency in comparison to a full theory, which takes into account the higher order idler generation as described in the previous section.  As discussed in Ref.~\onlinecite{li_efficient_2016}, quantitative agreement between theory and experiment can be achieved by including a larger basis of frequency modes, e.g., through adaptation of the mean-field Lugiato-Lefever equation (LLE) that has been used in modeling of microresonator frequency combs~\cite{coen_modeling_2013,chembo_spatiotemporal_2013}.

\vspace{-0.2in}

\section{QFC experimental setup}

\begin{center}
	\begin{figure*}[t]
		\begin{center}
			\includegraphics[width=0.9\linewidth]{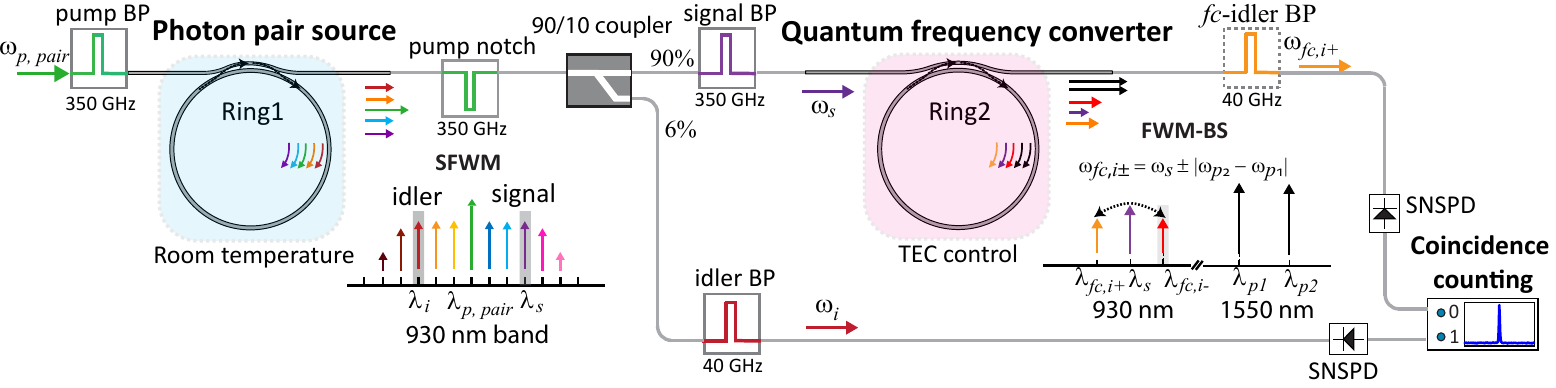}
			\caption{\textbf{Experimental schematic for QFC}. After selecting a specific set of signal and idler frequency bins from the photon pair source, the signal is combined with two 1550 nm pumps using a 930 nm/1550 nm WDM (not shown) and coupled to the frequency converter, with the idler directly transmitted to the detector. At the output of the frequency converter, light is separated into its respective frequency domains using another WDM (not shown) and the frequency-converted blue idler is filtered for coincidence measurement. BP, bandpass filter; SNSPD, superconducting single photon detector; and TEC, thermoelectric controller.}
			\label{EDF_QFC_schematic}
		\end{center}
	\end{figure*}
\end{center}

The measurements performed in Fig.~2 and Fig.~3 are based on the same experimental schematic shown in Fig.~\ref{EDF_QFC_schematic}, except that the two 1550 nm pumps for the frequency conversion are off in Fig.~2 (before conversion) and on in Fig.~3 (after conversion). The coupling efficiency between the lensed fiber and the photonic chip is 50~\% to 60~\% per facet by implementing inverse tapers at the ends of the waveguide. The overall transmission of the signal (idler) from the photon pair chip to the single photon detector is approximately 1.7~\% (0.72~\%), which is estimated by multiplying the transmission of various components in the optical path. For example, for the signal we have: out-coupling of the photon pair chip (50~\%), pump notch filter (60~\%), wavelength division multiplexer (WDM) to separate into two channels (90~\%), signal bandpass filter (50~\%), 1550/930 nm WDM to combine signal in the 930 nm band and two pumps in the 1550 nm band (80~\%), frequency converter chip (50~\% per facet), another 1550/930 nm WDM to separate light to the 1550 nm and 930 nm bands (80~\%), and superconducting nanowire single photon detectors (SNSPD, 80~\% detection efficiency), totaling to $1.73~\%$. Similarly, the overall transmission for the idler can be calculated ($0.72~\%$), noting that the 10 port of the WDM has an actual transmission of 6~\% instead of 10~\%. The photon count rate shown in the left $y$ axis in Fig.~2b and Figs.~3a,d,f corresponds to the detected photon flux. Error bars are one standard deviation values and stem from fluctuations in the detected flux.

\section{Noise analysis for quantum frequency conversion experiment}

In this section, we present a detailed analysis of noise in the quantum frequency conversion experiment. Specifically, we study the noise before frequency conversion (subsection V.A), how it changes after frequency conversion (subsection V.B), its impact on the CAR measurement (subsection V.C), and future directions for reducing noise (subsection V.D).

\subsection{Noise before frequency conversion}

\begin{center}
	\begin{figure*}[ht]
		\begin{center}
			\includegraphics[width=\linewidth]{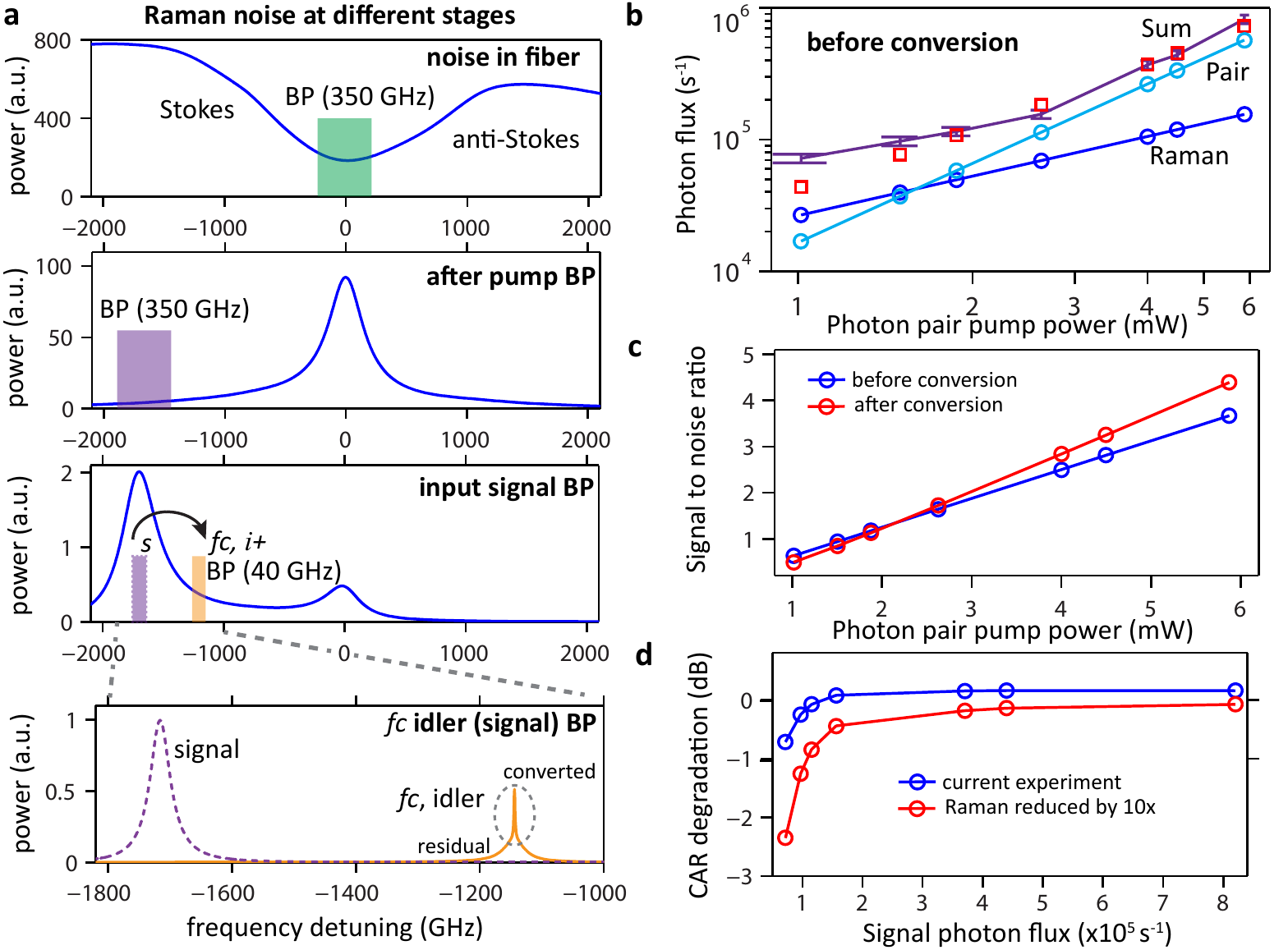}
			\caption{\textbf{Noise analysis for QFC}. \textbf{a}, Simulated Raman noise at various stages of the QFC experiment (counting from top to bottom): (1) in the optical fiber that connects the laser to the photon pair chip, with the $x$ axis being the frequency detuning relative to the pump frequency; (2) at the input of the photon pair chip after a 350 GHz bandwidth bandpass filter (1 nm in wavelength, see Fig.~\ref{EDF_QFC_schematic}) is spectrally aligned with the photon pair source signal frequency. The transmission of all the bandpass filters is taken to be 50~\%; (3) at the input of the frequency converter chip after using a 350 GHz bandwidth bandpass filter to select the signal; and (4) at the idler frequency after using a 40 GHz bandwidth bandpass filter to select the frequency converted blue idler, which consists of the frequency-converted part (sharp peak in the yellow solid curve) and the residual Raman noise from the input due to insufficient spectral suppression (wider base in the yellow solid curve). In comparison, we have also plotted the Raman noise at the signal frequency if we use a 40 GHz bandwidth bandpass filter to select the signal (violet dotted curve). \textbf{b}, Decomposing the detected photon counts from the photon pair source into Raman noise (blue circles) and photon pairs (cyan circles). Their sum (red squares) agrees with the experimental data reasonably well (violet line with error bars, also shown in Fig.~3f in the main text). \textbf{c} Estimated SNR values before and after frequency conversion based on a conversion efficiency of $25~\%$ and the various noise contributions described in the text. \textbf{d}, Degradation of CAR values for the QFC experiment shown in Fig.~3 in the main text (blue circles) as well as for a hypothetical scenario that the Raman noise on the input source is further suppressed by a factor of 10 through the use of narrower bandpass filters (red circles). The $x$ axis corresponds to the signal photon flux in the ``before conversion" case in \textbf{b}, which would be the on-chip signal photon flux if there is no coupling loss.}
			\label{FigS_noise_analysis}
		\end{center}
	\end{figure*}
\end{center}

As illustrated in Fig.~\ref{FigS_noise_analysis}a, there is a broadband background noise associated with the photon pairs generated in the first microring. This noise arises from Raman scattering induced by the pump during its propagation in the optical fiber that connects the laser to the photonic chip. Despite the use of multiple bandpass filters in the experiment, a complete suppression of the Raman noise is challenging due to the relatively small spectral separation between the selected photon pair and the pump. As a result, the signal-to-noise ratio (SNR) before conversion, which is defined as the ratio between the signal and the background noise at the input of the frequency converter, is limited. This SNR can be experimentally measured by comparing the photon counts in the relevant photon pair signal/idler frequency bins between when the pump is off-resonance with its relevant cavity mode (Raman noise only in the signal/idler bins) and the pump on-resonance case (Raman and photon pairs), and noting that Raman noise entering the chip is the same in these two cases. For example, the SNR for the pump power near 4 mW (corresponding to the spectrum shown in Fig.~2b in the main text) is estimated to be around 2.5, indicating that a significant portion of the detected photon flux is Raman noise ($29~\%$ at this power level). The presence of the Raman noise also explains the power dependence of the detected photon rates being different than quadratic. In fact, by decomposing the photon counts into the Raman noise and the photon pairs, with the former having a linear power dependence and the latter having a quadratic dependence, a reasonable agreement between theory and experiment has been achieved in Fig.~\ref{FigS_noise_analysis}b. We also plot the estimated SNRs before conversion in Fig.~\ref{FigS_noise_analysis}c (blue circles), which display a linear dependence on the pump power.

\subsection{Noise after frequency conversion}

The noise at the idler frequency after conversion is composed of several parts: (1) input noise at the signal frequency that gets frequency-converted to the idler channel; (2) residual input noise at the idler frequency (i.e., due to Raman); and (3) additional noise contributed by the frequency converter. Compared to the input noise, which is due to broadband Raman scattering and is spectrally distributed over the width of the bandpass filter used to select the frequency-converted idler channel (350~GHz bandwidth), the converted noise is significantly reduced since the conversion bandwidth (2 GHz) is much smaller than the bandpass filter bandwidth. In our experimental configuration, the total contribution of these two types of noise at the idler frequency amounts to approximately 20~\% of the input noise, while 25~\% of the photon pair signal is converted to the idler (without including coupling losses). On the other hand, the additional noise from the frequency converter is independent of the input noise and is solely determined by the two pumps in the 1550 nm band. This analysis suggests that the SNR after the conversion can be slightly improved if the input photon flux is much larger than the additional noise from the frequency converter (Fig.~\ref{FigS_noise_analysis}c, red circles). On the other hand, if the input photon flux is comparable to the frequency conversion noise, the SNR is expected to deteriorate after the frequency conversion. However, this would only happen when the on-chip pair generation rate is much smaller than $1.5\times 10^4$ pairs/s (3 fW) for the current device.

\subsection{CAR vs noise}
To study the impact of noise on the CAR measurement, we use a simplified model developed in Ref.~\cite{clemmen_continuous_2009} for the coincidence detection rate:
\begin{equation}
C_p=\gamma_e\eta_s\eta_i\tau_b, \label{Eq_Cp}
\end{equation}
where $\gamma_e$ is the on-chip pair generation rate, $\eta_s$ ($\eta_i$) represents the total loss for the signal (idler), and $\tau_b$ is the time-bin size. Similarly, the detection probability per unit time for the signal (or idler) is given by $p_{s,i}=\gamma_e\eta_{s,i}+dk_{s,i}$, where $dk_{s,i}$ denote the noise (including dark counts from detectors) associated with the signal and idler detection. If the photon flux is low enough, we can neglect the contribution from multi-photon generation, and the accidental coincidence detection rate is given by
\begin{equation}
C_a=(\gamma_e\eta_s+dk_s)(\gamma_e\eta_i+dk_i)\tau_b^2. \label{Eq_Ca}
\end{equation}
Using the definition of $\text{CAR}\equiv C_p/C_a$, we obtain the following expression after some simplification:
\begin{equation}
\text{CAR}=\frac{1}{\gamma_e\tau_b}\frac{1}{(1+1/\text{SNR}_{s})(1+1/\text{SNR}_{i})}, \label{Eq_CAR}
\end{equation}
where SNR$_{s,i}$ represent the SNRs for the signal and idler. The dependence of CAR on pump power, as shown in Fig.~\ref{FigS_photon_source}c, can be understood using Eq.~\ref{Eq_CAR}. For example, at high pump powers the term containing the SNR can be neglected (close to 1), and the CAR is inversely proportional to $\gamma_e$ (until multi-photon contribution becomes significant). On the other hand, for small enough pump powers, the term containing SNR would decrease much more rapidly than the reduced $\gamma_e$ (mainly due to fixed dark counts from the detectors), hence leading to the peak observed in the CAR measurement. Equation \ref{Eq_CAR} also explains the insensitivity of CAR to the additional coupling loss (Fig.~\ref{FigS_photon_source}c), simply because the SNR is kept the same during the coupling process.

In the QFC experiment, the CAR values before and after frequency conversion are compared. Since the same photon pair source idler is used in both CAR measurements, we only need to focus on the change in SNR for the signal photons when they get mapped to frequency-converted idler photons. The degradation of the CAR can therefore be obtained as:
\begin{equation}
\text{CAR}_\text{degradation} =\frac{1+ 1/\text{SNR}_{s,\text{before}}}{1+ 1/\text{SNR}_{s,\text{after}}}.
\end{equation}
A straightforward calculation for the QFC experiment performed in this work shows that the CAR is expected to stay almost the same as the before conversion case (blue circles in Fig.~\ref{FigS_noise_analysis}d), agreeing with our experimental data. In addition, we also have considered a hypothetical scenario in which the Raman noise is suppressed further by one order of magnitude. In this case, the Raman noise near the pump power of 1 mW becomes comparable to the additional noise from the frequency converter, leading to a more significant drop in the CAR values after conversion (red circles in Fig.~\ref{FigS_noise_analysis}d). Again, we want to stress here that the CAR values can be well-maintained if the signal photon flux is large enough.
\subsection{Steps for improved performance}

The above analysis indicates that the input Raman noise within the photon pair source currently limits the pair source CAR values to around 30, and is significantly larger than the noise added by the frequency converter. As a result, we do not observe any significant degradation in CAR in the quantum frequency conversion experiment. In this section, we consider steps that might improve the overall performance of the system.

We start with the frequency converter, noting that while its performance is suitable for the microresonator photon pair source we have developed in this work, reductions in added noise will be needed if either the source brightness or input noise from the source (e.g., the Raman contribution) are reduced (Fig.~\ref{FigS_noise_analysis}c-d). Low-loss, narrowband spectral filtering (at the GHz or few hundred MHz level) at the output of the frequency converter is an approach that is often used to reduce noise in QFC, for example, in several recent demonstrations of donwconversion to the telecommunications band use periodically-poled $\chi^{(2)}$ waveguides~\cite{albrecht_waveguide_2014,dreau_quantum_2018,walker_long-distance_2018,maring_quantum_2018}. In this work, we use a 40~GHz bandpass filter after the QFC chip.  However, we note that in contrast to all previous works on QFC, our device is resonator-based, and hence it naturally acts a spectral filter itself, with a bandwidth set by the conversion bandwidth (about 2~GHz in the experiment we present).  However, the configuration and extinction ratio of the converter is not optimized for filtering. Thus, an appealing approach would be to integrate an on-chip filter after the converter, and indeed, filters have been implemented in Si$_3$N$_4$ using both microring-based~\cite{barwicz_microring-resonator-based_2004} and grating-based~\cite{zhu_arbitrary_2016} geometries.  The bandwidth of this filter will depend on the source - a GHz level bandwidth is appropriate for the current pair sources we work with, but for much narrower-band photons, e.g., from a quantum emitter with a long lifetime, more aggressive narrowband filtering should be employed.  Considering that~$\approx10$~MHz linewidths have been achieved in Si$_3$N$_4$ microring resonators~\cite{ji_ultra-low-loss_2017}, there is significant scope for implementing narrower-band on-chip filters.

Ultimately, however, our frequency converter does add some amount of noise, which will be resonant with the frequency-converted photons and hence cannot be readily removed through spectral filtering. It seems unlikely to be due to Raman scattering in the Si$_3$N$_4$~\cite{dhakal_silicon-nitride_2014}, given that our signal and idler photons are separated by 130~THz (and on the anti-Stokes side) of the 1550~nm pump.  We tentatively hypothesize that the noise is due to fluorescence from the Si$_3$N$_4$ material, possibly due to nanocrystal formation during the film growth or annealing, which is known to be sensitive to specific growth parameters~\cite{basa_si_2007}. Future studies of the influence film growth and annealing conditions on added noise are needed to help elucidate such points.

Additional spectral filtering is also needed to improve the performance of our photon-pair source. In particular, this would aim to further suppress the broadband Raman noise that enters into the photon pair source and overlaps with its signal and idler frequency bins (as well as the converted frequency bin), as described in the previous section.  Similar to the output spectral filtering described above, such an input filter (i.e., placed before the photon pair source) could be implemented on-chip.

Finally, cryogenic temperatures are known to significantly reduce Raman noise in optical fibers~\cite{takesue_1.5-m_2005,lin_photon-pair_2007}, and is thus a straightforward path to improving the performance of the photon pair source (i.e., by limiting the noise entering the Si$_3$N$_4$ chip).  The influence of cryogenic temperatures on the performance of the Si$_3$N$_4$ devices themselves (e.g., in terms of the added noise produced by the frequency converter) is unknown, but worth exploring.  Furthermore, direct integration of SNSPDs with Si$_3$N$_4$ photonic circuits~\cite{schuck_nbtin_2013} would require cryogenic temperatures for operation.

\section{HOM interference experimental details}

\begin{center}
	\begin{figure*}[b]
		\begin{center}
			\includegraphics[width=0.9\linewidth]{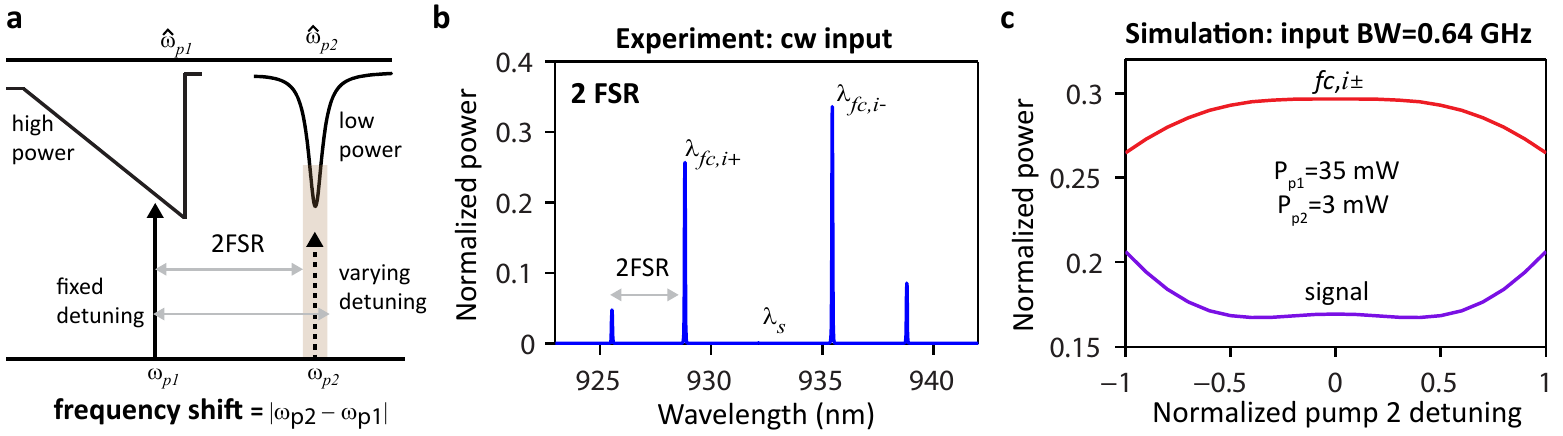}
			\caption{\textbf{Frequency shift tuning by pump lasers}. \textbf{a}, Experimental configuration for tuning the frequency shift between the signal and idlers in the 930 nm band by varying the frequency detuning of pump lasers in the 1550 nm band: pump 1 is set at a much higher power than pump 2, so that the pump 1 cavity resonance exhibits thermal bistability while the pump 2 cavity resonance resides in the linear regime. By fixing the detuning of pump 1 while varying the detuning of pump 2, we can thus tune the spectral translation range of our frequency converter without modify its thermal dynamics. \textbf{b}, Experimental frequency conversion spectrum following this approach, with the pump frequency separation of about 2 FSRs ($\approx 1.14$ THz), and using a cw 930~nm band input signal. The optical spectrum has been normalized by the signal input power, showing a conversion efficiency near $34\%$ for the red idler. \textbf{c}, Simulated conversion efficiencies for a pulsed signal input (bandwidth around 0.64 GHz) as a function of the pump 2 laser detuning. Here we assume the signal detuning is fixed at zero, and the frequency shift is equal to 2 FSRs in the 930 nm band when the pump 2 laser is set at zero detuning. The $x$ axis is normalized by half of the cavity linewidth (0.5~GHz).}
			\label{EDF_pump_tuning}
		\end{center}
	\end{figure*}
\end{center}

In the HOM interference experiment carried out in Fig.~4 we employ the same frequency converter that has been used for the demonstration of quantum frequency conversion in Fig.~3, but replace the photon pair source by another microring that has approximately the same FSR as the frequency converter (by simply choosing a different ring width). Tuning of the frequency converter's spectral translation range to ensure the precise spectral matching needed for HOM interference is achieved by varying the frequency difference of the two pump lasers that drive the frequency conversion process. While this pump frequency difference sets the spectral translation range, the efficiency of the process depends on the pump powers coupled into the resonator, which in turn depend on the detuning of each pump with respect to its cavity mode. When using multiple strong pumps, a challenge is in simultaneously setting the detuning of both pumps at desired values, as adjustment of one pump frequency influences the detuning of the other, due to the thermo-optic effect (i.e., changes in the in-coupled pump power influence the location of all of the cavity modes). Typically, an iterative procedure is required, and this is what has been employed in all experiments in which the pump frequencies are not varied.

In the HOM interference experiments, we need to actively vary the pump frequency separation. To avoid having to implement the iterative approach for every desired frequency separation values, we instead take advantage of the fact that FWM-BS efficiency depends on the product of the two pump powers. This means that low power in one pump can be largely compensated by increasing the power in the second pump. The approach is illustrated in Fig.~\ref{EDF_pump_tuning}a. We set one pump (pump 2) at a low enough power so that its cavity resonance stays in the linear regime, whereas the other pump (pump 1) is set at a high enough power to provide the desired conversion efficiency. In this configuration, the thermal dynamics of the microresonator is dominated by the strong pump (pump 1 with fixed laser detuning) and is almost independent of the detuning of the weak pump 2. This allows us to freely adjust the detuning of pump 2 (either blue- or red-detuned with respect to its cavity resonance) without affecting the detuning of pump 1 with respect to its resonance. Figure~\ref{EDF_pump_tuning}b shows the output converted spectrum of a classical cw laser input, in the case where the pump frequency separation is about 2 FSRs, and indicates that conversion efficiency $>30~\%$ can be achieved. Figure~\ref{EDF_pump_tuning}c presents the results of a simulation (this time using a pulsed input source with linewidth of 640~MHz), showing that the peak conversion efficiency is largely maintained as the frequency of pump 2 (the low power pump) is varied on the order of the cavity linewidth.

\begin{center}
	\begin{figure*}[t]
		\begin{center}
			\includegraphics[width=0.9\linewidth]{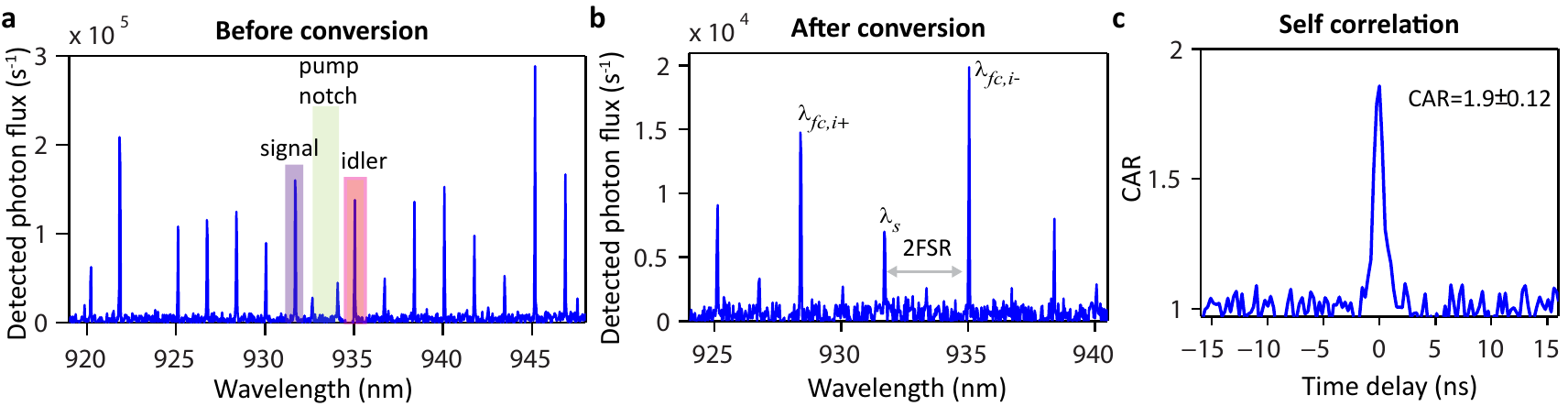}
			\caption{\textbf{Additional experimental data for two photon interference}. \textbf{a}, Spectrum of the photon pair source used in the two photon interference experiment in Fig.~5. \textbf{b}, Spectrum after frequency conversion. \textbf{c}, Self-correlation measurement of the pair source idler (or frequency-converted idler) using the schematic shown in Fig.~4, with one input arm of the 50/50 coupler left open.}
			\label{EDF_self_correlation}
		\end{center}
	\end{figure*}
\end{center}

Figure~\ref{EDF_self_correlation}a shows the spectrum of the photon pair source used in the HOM interference experiment, where the first-order pair is chosen as the signal and idler. The spectrum after conversion is plotted in Fig.~\ref{EDF_self_correlation}b, and the conversion efficiency of the red idler is estimated to be around $23\%\pm3\%$. Moreover, we have measured the self-correlation of the idler (or the frequency-converted red idler) using the experimental schematic shown in Fig.~5, by leaving one input arm of the 50/50 coupler open. The measured CAR value (Fig.~\ref{EDF_self_correlation}c) around 2 meets the expectation for such probabilistic photon sources.

\section{Hong-Ou-Mandel interference: theoretical model}
Finally, in this section we provide a brief calculation of the coincidence rate in the Hong-Ou-Mandel interference experiment. Even though in our experiment it is the interference between the frequency-converted red idler (from the signal) and the original idler, here for convenience we call them signal and idler. Assuming 1, 2 are the two input ports and 3, 4 are the two output ports, the fields at the output of a 50/50 coupler (or beam splitter) can be expressed as
\begin{align}
\hat{E}_3^{+}(t) & = N_3\left(\hat{a}_1(\tau_1) e^{-i\omega_1(t-\tau_1)} + i \hat{a}_2 (\tau_2)e^{-i\omega_2 (t-\tau_2)}\right), \label{Eq_E3}\\
\hat{E}_4^{+}(t) & = N_4\left(\hat{a}_2(\tau_2) e^{-i\omega_2(t-\tau_2)} + i \hat{a}_1 (\tau_1)e^{-i\omega_1 (t-\tau_1)}\right), \label{Eq_E4}
\end{align}
where $\hat{a}_n (n=1,2)$ is the destruction operation for the $i_{th}$ input, $\omega_n$ is the corresponding center frequency, $\tau_n$ is the traveling time for the $n_{th}$ photon to reach detectors, and $N_3$ and $N_4$ are normalization parameters. The coincidence rate is given by
\begin{equation}
P_{34}(t, t+\tau)=\bra{\mathbf{1}}_s\bra{\mathbf{1}}_i \hat{E}_3^{-}(t)\hat{E}_4^{-}(t+\tau)\hat{E}_4^{+}(t+\tau) \hat{E}_3^{+}(t)\ket{\mathbf{1}}_i\ket{\mathbf{1}}_s. \label{Eq_P34}
\end{equation}
Substituting Eqs.~\ref{Eq_E3} and \ref{Eq_E4} into Eq.~\ref{Eq_P34} results:
\begin{align}
P_{34}(t, t+\tau)=\left|N_3N_4\right|^2 \Big\{&\bra{\mathbf{1}}_s \hat{a}_1^{+}(\tau_1) \hat{a}_1(\tau_1) \ket{\mathbf{1}}_s \cdot \bra{\mathbf{1}}_i \hat{a}_2^{+}(\tau_2+ \tau) \hat{a}_2(\tau_2+\tau) \ket{\mathbf{1}}_i   \notag \\
+ &\bra{\mathbf{1}}_s \hat{a}_1^{+}(\tau_1+\tau) \hat{a}_1(\tau_1+ \tau) \ket{\mathbf{1}}_s \cdot \bra{\mathbf{1}}_i \hat{a}_2^{+}(\tau_2) \hat{a}_2(\tau_2) \ket{\mathbf{1}}_i   \notag  \\
- & \bra{\mathbf{1}}_i \hat{a}_2^{+}(\tau_2+\tau) \hat{a}_1(\tau_1+ \tau) \ket{\mathbf{1}}_s \cdot \bra{\mathbf{1}}_s \hat{a}_1^{+}(\tau_1) \hat{a}_2(\tau_2) \ket{\mathbf{1}}_i e^{-i\Delta\omega_{12}\tau}   \notag \\
- & \bra{\mathbf{1}}_i \hat{a}_2^{+}(\tau_2) \hat{a}_1(\tau_1) \ket{\mathbf{1}}_s \cdot \bra{\mathbf{1}}_s \hat{a}_1^{+}(\tau_1+\tau) \hat{a}_2(\tau_2+\tau) \ket{\bm{1}}_i e^{i\Delta\omega_{12}\tau}  \Big\}, \label{Eq_P34_2}
\end{align}
where $\Delta\omega_{12}\equiv \omega_1 -\omega_2$ is the frequency difference of the two input photons.

Equation \ref{Eq_P34_2} has four components essentially describing the temporal correlations between the signal and idler photons: The first two terms describe the coincidence rate as if they are directly detected, and the last two terms describe the interference between them. These expressions can be computed based on the fact that, the signal and idler photons are created at the same time inside the microring, but they can couple out to the waveguide at different times with probability exponentially decaying with their time separation (i.e., $\exp(-\gamma|t_\text{sep}|)$ with $\gamma$ being the photon decay rate). As a result, we have
\begin{equation}
P_{34}(t, t+\tau)= \text{const.}\times\left\{e^{-\gamma|\tau-\Delta t_{12}|} + e^{-\gamma|\tau+\Delta t_{12}|} -2e^{-\gamma|\tau|}\cos(\Delta \omega_{12}\tau) \left(\hat{\bm{e}}_s\cdot\hat{\bm{e}_i}\right)^2 \right\}, \label{Eq_P34_3}
\end{equation}
where $\Delta t_{12}\equiv \tau_1-\tau_2$ is the arrival time difference between the two photons, and $\hat{\bm{e}}_s$ ($\hat{\bm{e}}_i$) is the unit polarization vector for the signal (idler) photon.

The above equation allows us to obtain theoretical predictions for the two-photon interference experiment performed in Fig.~4 in the main text. For example, for the case that the two photons arrive at the 50/50 coupler at the same time ($\Delta t_{12}=0$) with different frequencies, we have
\begin{align}
P_{34}^{\perp}(t, t+\tau) &= \text{const.} \times  2 e^{-\gamma |\tau|} \ \ \ (\text{orthogonal polarization}), \\
P_{34}^{\parallel}(t, t+ \tau) &= \text{const.} \times 2 e^{-\gamma |\tau|} \left(1- \cos\left(\Delta\omega_{12}\tau\right)\right) \ \ \ (\text{parallel polarization}).
\end{align}
Similarly, for photons arriving at the 50/50 coupler with a delay with respect to each other but with the same frequency, we have
\begin{align}
P_{34}^{\perp}(t, t+\tau) &= \text{const.} \times \left(e^{-\gamma|\tau-\Delta t_{12}|} + e^{-\gamma|\tau+\Delta t_{12}|}  \right)\ \ \ (\text{orthogonal polarization}), \\
P_{34}^{\parallel}(t, t+ \tau) &= \text{const.} \times \left(e^{-\gamma|\tau-\Delta t_{12}|} + e^{-\gamma|\tau+\Delta t_{12}|} -2e^{-\gamma|\tau|} \right) \ \ \ (\text{parallel polarization}).
\end{align}
The visibility plotted in Fig.~4d in the main text is defined as the contrast of coincidence peaks between the parallel and orthogonal polarizations:
\begin{equation}
\text{Visiblity}\equiv \frac{P_{34,\text{max}}^{\perp} -P_{34,\text{max}}^{\parallel}}{P_{34,\text{max}}^{\perp} +P_{34,\text{max}}^{\parallel}}.
\end{equation}
							
\section{Comparison of electro-optic phase modulator and FWM-BS approaches}

\begin{center}
	\begin{figure}[b]
		\begin{center}
			\includegraphics[width=0.55\linewidth]{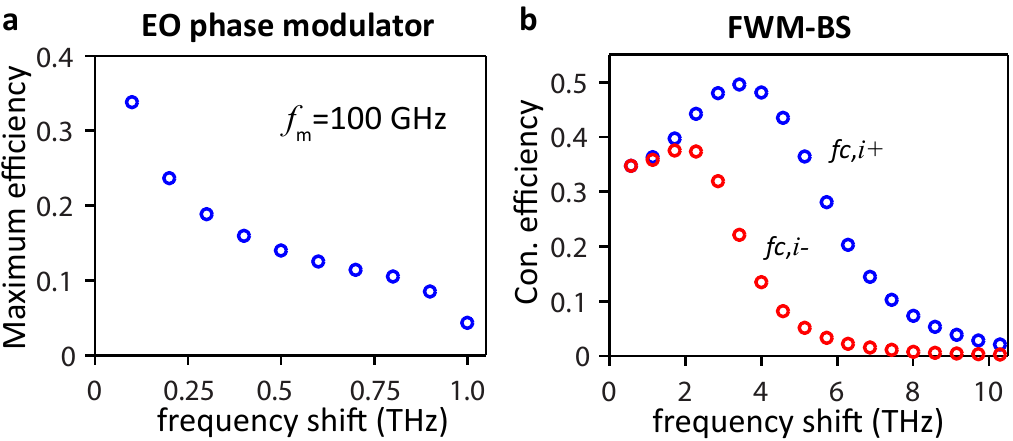}
			\caption{\textbf{Comparison between the phase modulator and the FWM-BS process}. \textbf{a}, Maximum possible efficiency for different sidebands in an electric-optic phase modulator with $f_m=100$ GHz. \textbf{b}, Simulated conversion efficiency for a 40 $\mu$m radius Si$_3$N$_4$ microring with matched FSR between the 1550 nm and 930 nm bands. In the simulation, we have used $D_2/2\pi=-23.3$ MHz and $D_3/2\pi=-1.07$ MHz for the 1550 nm band, and $D_2/2\pi=-16.5$ MHz and $D_3/2\pi=0.78$ MHz for the 930 nm.}
			\label{EDF_PM_vs_FWMBS}
		\end{center}
	\end{figure}
\end{center}

Here we briefly compare the performance of an electro-optic (EO) phase modulator and the FWM-BS process for the frequency translation of optical signals. First, an optical carrier signal going through an EO phase modulator can be expressed in the following form:
\begin{equation}
E_m(t)=E_0 \sum_{n} J_n(\phi_0)\cos\left[(2\pi f_0 + n2\pi f_m)\right],  \label{Eq_PM}
\end{equation}

where $E_0$ is a constant, $J_n(\phi_0)$ denotes the Bessel function of the first kind and of order $n$ ($n$ is an integer and $\phi_0$ is the modulation depth), $f_0$ is the carrier frequency, and $f_m$ is the modulation frequency. According to Eq.~\ref{Eq_PM}, the normalized power at the $n_\text{th}$ sideband is simply given by $J_n^2(\phi_0)$, whose peak value by varying $\phi_0$ corresponds to the maximum possible efficiency for a frequency translation of $nf_m$ (Fig.~\ref{EDF_PM_vs_FWMBS}a). While the conversion efficiency for the first few sidebands can be decent ($\approx 10\%-35\%$), it drops quickly below $10\%$ at higher-order sidebands. Typically, $f_m$ is limited to 100 GHz by the available electronic bandwidth, therefore rendering efficient frequency translation beyond 1 THz difficult.

On the other hand, in the FWM-BS process, the conversion efficiency is determined by the frequency matching condition and the process can be efficient over a broad range. For example, Fig.~\ref{EDF_PM_vs_FWMBS}b plots the expected conversion efficiency for the intraband frequency conversion when there is $D_{1}$ matching between the 1550 nm and 930 nm bands (i.e., as in our demonstrated devices). The blue idler maintains a high conversion efficiency ($>30\%$) for a more than 5-THz-wide spectral window. A reduction in second order ($D_{2}$) and higher order dispersion terms can enable wider spectral translation ranges to be achieved. We can further improve the efficiency by reducing the intrinsic propagation loss of the microresonator, increasing the level of over-coupling, and engineering the dispersion to enhance one idler while suppressing the other.


\begin{thebibliography}{10}
\newcommand{\enquote}[1]{``#1''}
\expandafter\ifx\csname url\endcsname\relax
  \def\url#1{{#1}}\fi
\expandafter\ifx\csname urlprefix\endcsname\relax\def\urlprefix{}\fi

\bibitem{lukens_frequency-encoded_2017}
J.~M. Lukens and P.~Lougovski, \enquote{Frequency-encoded photonic qubits for
  scalable quantum information processing,} Optica {\bf 4}, 8--16 (2017).

\bibitem{kues_-chip_2017}
M.~Kues, C.~Reimer, P.~Roztocki, L.~R. Cortés, S.~Sciara, B.~Wetzel, Y.~Zhang,
  A.~Cino, S.~T. Chu, B.~E. Little, D.~J. Moss, L.~Caspani, J.~Azaña, and
  R.~Morandotti, \enquote{On-chip generation of high-dimensional entangled
  quantum states and their coherent control,} Nature {\bf 546}, 622--626
  (2017).

\bibitem{imany_50-ghz-spaced_2018}
P.~Imany, J.~A. Jaramillo-Villegas, O.~D. Odele, K.~Han, D.~E. Leaird, J.~M.
  Lukens, P.~Lougovski, M.~Qi, and A.~M. Weiner, \enquote{50-{GHz}-spaced comb
  of high-dimensional frequency-bin entangled photons from an on-chip silicon
  nitride microresonator,} Optics Express {\bf 26}, 1825 (2018).

\bibitem{caspani_integrated_2017}
L.~Caspani, C.~Xiong, B.~J. Eggleton, D.~Bajoni, M.~Liscidini, M.~Galli,
  R.~Morandotti, and D.~J. Moss, \enquote{Integrated sources of photon quantum
  states based on nonlinear optics,} Light: Science \& Applications {\bf 6},
  e17\,100 (2017).

\bibitem{mckinstrie_translation_2005}
C.~McKinstrie, J.~Harvey, S.~Radic, and M.~Raymer, \enquote{Translation of
  quantum states by four-wave mixing in fibers,} Optics Express {\bf 13},
  9131--9142 (2005).

\bibitem{li_efficient_2016}
Q.~Li, M.~Davan\c{c}o, and K.~Srinivasan, \enquote{Efficient and low-noise
  single-photon-level frequency conversion interfaces using silicon
  nanophotonics,} Nature Photonics {\bf 10}, 406--414 (2016).

\bibitem{singh_quantum_2019}
A.~Singh, Q.~Li, S.~Liu, Y.~Yu, X.~Lu, C.~Schneider, S.~Höfling, J.~Lawall,
  V.~Verma, R.~Mirin, S.~W. Nam, J.~Liu, and K.~Srinivasan, \enquote{Quantum
  frequency conversion of a quantum dot single-photon source on a nanophotonic
  chip,} Optica {\bf 6}, 563 (2019).

\bibitem{olislager_frequency-bin_2010}
L.~Olislager, J.~Cussey, A.~T. Nguyen, P.~Emplit, S.~Massar, J.-M. Merolla, and
  K.~P. Huy, \enquote{Frequency-bin entangled photons,} Physical Review A {\bf
  82}, 013\,804 (2010).

\bibitem{legero_quantum_2004}
T.~Legero, T.~Wilk, M.~Hennrich, G.~Rempe, and A.~Kuhn, \enquote{Quantum {Beat}
  of {Two} {Single} {Photons},} Physical Review Letters {\bf 93}, 070\,503
  (2004).

\bibitem{kumar_quantum_1990}
P.~Kumar, \enquote{Quantum frequency conversion,} Optics Letters {\bf 15},
  1476--1478 (1990).

\bibitem{raymer_manipulating_2012}
M.~G. Raymer and K.~Srinivasan, \enquote{Manipulating the color and shape of
  single photons,} Physics Today {\bf 65}, 32--37 (2012).

\bibitem{tanzilli_photonic_2005}
S.~Tanzilli, W.~Tittel, M.~Halder, O.~Alibart, P.~Baldi, N.~Gisin, and
  H.~Zbinden, \enquote{A photonic quantum information interface,} Nature {\bf
  437}, 116--120 (2005).

\bibitem{rakher_quantum_2010}
M.~T. Rakher, L.~Ma, O.~Slattery, X.~Tang, and K.~Srinivasan, \enquote{Quantum
  transduction of telecommunications-band single photons from a quantum dot by
  frequency upconversion,} Nature Photonics {\bf 4}, 786--791 (2010).

\bibitem{mcguinness_quantum_2010}
H.~J. McGuinness, M.~G. Raymer, C.~J. McKinstrie, and S.~Radic,
  \enquote{Quantum {Frequency} {Translation} of {Single}-{Photon} {States} in a
  {Photonic} {Crystal} {Fiber},} Physical Review Letters {\bf 105}, 093\,604
  (2010).

\bibitem{zaske_visible--telecom_2012}
S.~Zaske, A.~Lenhard, C.~A. Keßler, J.~Kettler, C.~Hepp, C.~Arend,
  R.~Albrecht, W.-M. Schulz, M.~Jetter, P.~Michler, and C.~Becher,
  \enquote{Visible-to-{Telecom} {Quantum} {Frequency} {Conversion} of {Light}
  from a {Single} {Quantum} {Emitter},} Physical Review Letters {\bf 109},
  147\,404 (2012).

\bibitem{ates_two-photon_2012}
S.~Ates, I.~Agha, A.~Gulinatti, I.~Rech, M.~T. Rakher, A.~Badolato, and
  K.~Srinivasan, \enquote{Two-{Photon} {Interference} {Using}
  {Background}-{Free} {Quantum} {Frequency} {Conversion} of {Single} {Photons}
  {Emitted} by an {InAs} {Quantum} {Dot},} Physical Review Letters {\bf 109},
  147\,405 (2012).

\bibitem{albrecht_waveguide_2014}
B.~Albrecht, P.~Farrera, X.~Fernandez-Gonzalvo, M.~Cristiani, and
  H.~de~Riedmatten, \enquote{A waveguide frequency converter connecting
  rubidium-based quantum memories to the telecom {C}-band,} Nature
  Communications {\bf 5}, 3376 (2014).

\bibitem{clemmen_ramsey_2016}
S.~Clemmen, A.~Farsi, S.~Ramelow, and A.~L. Gaeta, \enquote{Ramsey
  {Interference} with {Single} {Photons},} Physical Review Letters {\bf 117},
  223\,601 (2016).

\bibitem{wright_spectral_2017}
L.~J. Wright, M.~Karpiński, C.~Söller, and B.~J. Smith, \enquote{Spectral
  {Shearing} of {Quantum} {Light} {Pulses} by {Electro}-{Optic} {Phase}
  {Modulation},} Physical Review Letters {\bf 118}, 023\,601 (2017).

\bibitem{walker_long-distance_2018}
T.~Walker, K.~Miyanishi, R.~Ikuta, H.~Takahashi, S.~Vartabi~Kashanian,
  Y.~Tsujimoto, K.~Hayasaka, T.~Yamamoto, N.~Imoto, and M.~Keller,
  \enquote{Long-{Distance} {Single} {Photon} {Transmission} from a {Trapped}
  {Ion} via {Quantum} {Frequency} {Conversion},} Physical Review Letters {\bf
  120}, 203\,601 (2018).

\bibitem{maring_quantum_2018}
N.~Maring, D.~Lago-Rivera, A.~Lenhard, G.~Heinze, and H.~de~Riedmatten,
  \enquote{Quantum frequency conversion of memory-compatible single photons
  from 606 nm to the telecom {C}-band,} Optica {\bf 5}, 507--513 (2018).

\bibitem{dreau_quantum_2018}
A.~Dreau, A.~Tcheborateva, A.~E. Mahdaoui, C.~Bonato, and R.~Hanson,
  \enquote{Quantum frequency conversion to telecom of single photons from a
  nitrogen-vacancy center in diamond,} Physical Review Applied {\bf 9},
  064\,031 (2018).

\bibitem{siverns_neutral_2018}
J.~D. Siverns, J.~Hannegan, and Q.~Quraishi, \enquote{Neutral atom wavelength
  compatible 780 nm single photons from a trapped ion via quantum frequency
  conversion,} arXiv:1801.01193 [physics, physics:quant-ph]  (2018).

\bibitem{moss_new_2013}
D.~J. Moss, R.~Morandotti, A.~L. Gaeta, and M.~Lipson, \enquote{New
  {CMOS}-compatible platforms based on silicon nitride and {Hydex} for
  nonlinear optics,} Nature Photonics {\bf 7}, 597--607 (2013).

\bibitem{barwicz_microring-resonator-based_2004}
T.~Barwicz, M.~A. Popovic, P.~T. Rakich, M.~R. Watts, H.~A. Haus, E.~P. Ippen,
  and H.~I. Smith, \enquote{Microring-resonator-based add-drop filters in
  {SiN}: fabrication and analysis,} Optics Express {\bf 12}, 1437--1442 (2004).

\bibitem{xiong_compact_2015}
C.~Xiong, X.~Zhang, A.~Mahendra, J.~He, D.-Y. Choi, C.~J. Chae, D.~Marpaung,
  A.~Leinse, R.~G. Heideman, M.~Hoekman, C.~G.~H. Roeloffzen, R.~M.
  Oldenbeuving, P.~W.~L. van Dijk, C.~Taddei, P.~H.~W. Leong, and B.~J.
  Eggleton, \enquote{Compact and reconfigurable silicon nitride time-bin
  entanglement circuit,} Optica {\bf 2}, 724--727 (2015).

\bibitem{schuck_nbtin_2013}
C.~Schuck, W.~H.~P. Pernice, and H.~X. Tang, \enquote{{NbTiN} superconducting
  nanowire detectors for visible and telecom wavelengths single photon counting
  on {Si}3N4 photonic circuits,} Applied Physics Letters {\bf 102}, 051\,101
  (2013).

\bibitem{xue_thermal_2016}
X.~Xue, Y.~Xuan, C.~Wang, P.-H. Wang, Y.~Liu, B.~Niu, D.~E. Leaird, M.~Qi, and
  A.~M. Weiner, \enquote{Thermal tuning of {Kerr} frequency combs in silicon
  nitride microring resonators,} Optics Express {\bf 24}, 687 (2016).

\bibitem{hong_measurement_1987}
C.~K. Hong, Z.~Y. Ou, and L.~Mandel, \enquote{Measurement of subpicosecond time
  intervals between two photons by interference,} Phys. Rev. Lett. {\bf 59},
  2044--2046 (1987).

\bibitem{legero_characterization_2006}
T.~Legero, T.~Wilk, A.~Kuhn, and G.~Rempe, \enquote{Characterization of
  {Single} {Photons} {Using} {Two}-{Photon} {Interference},} in {\em Advances
  {In} {Atomic}, {Molecular}, and {Optical} {Physics}\/} (Elsevier, 2006),
  Vol.~53, pp. 253--289.

\bibitem{lu_electro-optic_2018}
H.-H. Lu, J.~M. Lukens, N.~A. Peters, O.~D. Odele, D.~E. Leaird, A.~M. Weiner,
  and P.~Lougovski, \enquote{Electro-{Optic} {Frequency} {Beam} {Splitters} and
  {Tritters} for {High}-{Fidelity} {Photonic} {Quantum} {Information}
  {Processing},} Physical Review Letters {\bf 120}, 030\,502 (2018).

\bibitem{coen_modeling_2013}
S.~Coen, H.~G. Randle, T.~Sylvestre, and M.~Erkintalo, \enquote{Modeling of
  octave-spanning {Kerr} frequency combs using a generalized mean-field
  {Lugiato}–{Lefever} model,} Optics Letters {\bf 38}, 37--39 (2013).

\bibitem{chembo_spatiotemporal_2013}
Y.~K. Chembo and C.~R. Menyuk, \enquote{Spatiotemporal {Lugiato}-{Lefever}
  formalism for {Kerr}-comb generation in whispering-gallery-mode resonators,}
  Physical Review A {\bf 87} (2013).

\bibitem{clemmen_continuous_2009}
S.~Clemmen, K.~P. Huy, W.~Bogaerts, R.~G. Baets, P.~Emplit, and S.~Massar,
  \enquote{Continuous wave photon pair generation in silicon-on-insulator
  waveguides and ring resonators,} Optics Express {\bf 17}, 16\,558 (2009).

\bibitem{zhu_arbitrary_2016}
T.~Zhu, Y.~Hu, P.~Gatkine, S.~Veilleux, J.~Bland-Hawthorn, and M.~Dagenais,
  \enquote{Arbitrary on-chip optical filter using complex waveguide {Bragg}
  gratings,} Applied Physics Letters {\bf 108}, 101\,104 (2016).

\bibitem{ji_ultra-low-loss_2017}
X.~Ji, F.~A.~S. Barbosa, S.~P. Roberts, A.~Dutt, J.~Cardenas, Y.~Okawachi,
  A.~Bryant, A.~L. Gaeta, and M.~Lipson, \enquote{Ultra-low-loss on-chip
  resonators with sub-milliwatt parametric oscillation threshold,} Optica {\bf
  4}, 619--624 (2017).

\bibitem{dhakal_silicon-nitride_2014}
A.~Dhakal, P.~Wuytens, F.~Peyskens, A.~Z. Subramanian, N.~Le~Thomas, and
  R.~Baets, \enquote{Silicon-nitride waveguides for on-chip {Raman}
  spectroscopy,} Proceedings of SPIE {\bf 9141}, 91\,411C (2014).

\bibitem{basa_si_2007}
P.~Basa, P.~Petrik, M.~Fried, L.~Dobos, B.~Pécz, and L.~Tóth, \enquote{Si
  nanocrystals in silicon nitride: {An} ellipsometric study using parametric
  semiconductor models,} Physica E: Low-dimensional Systems and Nanostructures
  {\bf 38}, 76--79 (2007).

\bibitem{takesue_1.5-m_2005}
H.~Takesue and K.~Inoue, \enquote{1.5-µm band quantum-correlated photon pair
  generation in dispersion-shifted fiber: suppression of noise photons by
  cooling fiber,} Optics Express {\bf 13}, 7832--7839 (2005).

\bibitem{lin_photon-pair_2007}
Q.~Lin, F.~Yaman, and G.~P. Agrawal, \enquote{Photon-pair generation in optical
  fibers through four-wave mixing: {Role} of {Raman} scattering and pump
  polarization,} Physical Review A {\bf 75}, 023\,803 (2007).

\end{thebibliography}

\end{document}